\documentclass[longauth]{aa_mod} 

\usepackage{graphicx}
\usepackage{txfonts}
\usepackage[]{hyperref}
\usepackage[switch]{lineno}

\usepackage{xcolor}

\def\SiII{Si\,{\sc ii}}
\def\gps{$g_{\rm P1}$}
\def\rps{$r_{\rm P1}$}
\def\ips{$i_{\rm P1}$}
\def\zps{$z_{\rm P1}$}
\def\yps{$y_{\rm P1}$}

\begin{document}

   \title{Search for transient optical counterparts to high-energy IceCube neutrinos with Pan-STARRS1}

   \author{
{\small
\textbf{Pan-STARRS:}
E.~Kankare\inst{\ref{Belfast},\ref{Turku}}\thanks{E-mail: e.kankare@qub.ac.uk (EK)}
\and M.~Huber\inst{\ref{Honolulu}}
\and S.~J.~Smartt\inst{\ref{Belfast}}
\and K.~Chambers\inst{\ref{Honolulu}}
\and K.~W.~Smith\inst{\ref{Belfast}}
\and O.~McBrien\inst{\ref{Belfast}}
\and T.-W. Chen\inst{\ref{MPE}}
\and H. Flewelling\inst{\ref{Honolulu}}
\and T. Lowe\inst{\ref{Honolulu}} 
\and E. Magnier\inst{\ref{Honolulu}} 
\and A. Schultz\inst{\ref{Honolulu}} 
\and C. Waters\inst{\ref{Honolulu}} 
\and R. J. Wainscoat\inst{\ref{Honolulu}}
\and M. Willman\inst{\ref{Honolulu}}
\and D. Wright\inst{\ref{minnesota}}
\and D. Young\inst{\ref{Belfast}}\\
\vspace{0.15cm}
\textbf{IceCube collaboration\thanks{E-mail: analysis@icecube.wisc.edu}: }
M. G. Aartsen\inst{\ref{christchurch}}
\and M. Ackermann\inst{\ref{zeuthen}}
\and J. Adams\inst{\ref{christchurch}}
\and J. A. Aguilar\inst{\ref{brusselslibre}}
\and M. Ahlers\inst{\ref{copenhagen}}
\and M. Ahrens\inst{\ref{stockholmokc}}
\and C. Alispach\inst{\ref{geneva}}
\and D. Altmann\inst{\ref{erlangen}}
\and K. Andeen\inst{\ref{marquette}}
\and T. Anderson\inst{\ref{pennphys}}
\and I. Ansseau\inst{\ref{brusselslibre}}
\and G. Anton\inst{\ref{erlangen}}
\and C. Arg\"uelles\inst{\ref{mit}}
\and J. Auffenberg\inst{\ref{aachen}}
\and S. Axani\inst{\ref{mit}}
\and P. Backes\inst{\ref{aachen}}
\and H. Bagherpour\inst{\ref{christchurch}}
\and X. Bai\inst{\ref{southdakota}}
\and A. Barbano\inst{\ref{geneva}}
\and S. W. Barwick\inst{\ref{irvine}}
\and V. Baum\inst{\ref{mainz}}
\and R. Bay\inst{\ref{berkeley}}
\and J. J. Beatty\inst{\ref{ohio},\ref{ohioastro}}
\and K.-H. Becker\inst{\ref{wuppertal}}
\and J. Becker Tjus\inst{\ref{bochum}}
\and S. BenZvi\inst{\ref{rochester}}
\and D. Berley\inst{\ref{maryland}}
\and E. Bernardini\inst{\ref{zeuthen}}
\and D. Z. Besson\inst{\ref{kansas}}
\and G. Binder\inst{\ref{lbnl},\ref{berkeley}}
\and D. Bindig\inst{\ref{wuppertal}}
\and E. Blaufuss\inst{\ref{maryland}}
\and S. Blot\inst{\ref{zeuthen}}
\and C. Bohm\inst{\ref{stockholmokc}}
\and M. B\"orner\inst{\ref{dortmund}}
\and S. B\"oser\inst{\ref{mainz}}
\and O. Botner\inst{\ref{uppsala}}
\and E. Bourbeau\inst{\ref{copenhagen}}
\and J. Bourbeau\inst{\ref{madisonpac}}
\and F. Bradascio\inst{\ref{zeuthen}}
\and J. Braun\inst{\ref{madisonpac}}
\and H.-P. Bretz\inst{\ref{zeuthen}}
\and S. Bron\inst{\ref{geneva}}
\and J. Brostean-Kaiser\inst{\ref{zeuthen}}
\and A. Burgman\inst{\ref{uppsala}}
\and R. S. Busse\inst{\ref{madisonpac}}
\and T. Carver\inst{\ref{geneva}}
\and C. Chen\inst{\ref{georgia}}
\and E. Cheung\inst{\ref{maryland}}
\and D. Chirkin\inst{\ref{madisonpac}}
\and K. Clark\inst{\ref{snolab}}
\and L. Classen\inst{\ref{munster}}
\and G. H. Collin\inst{\ref{mit}}
\and J. M. Conrad\inst{\ref{mit}}
\and P. Coppin\inst{\ref{brusselsvrije}}
\and P. Correa\inst{\ref{brusselsvrije}}
\and D. F. Cowen\inst{\ref{pennphys},\ref{pennastro}}
\and R. Cross\inst{\ref{rochester}}
\and P. Dave\inst{\ref{georgia}}
\and J. P. A. M. de Andr\'e\inst{\ref{michigan}}
\and C. De Clercq\inst{\ref{brusselsvrije}}
\and J. J. DeLaunay\inst{\ref{pennphys}}
\and H. Dembinski\inst{\ref{bartol}}
\and K. Deoskar\inst{\ref{stockholmokc}}
\and S. De Ridder\inst{\ref{gent}}
\and P. Desiati\inst{\ref{madisonpac}}
\and K. D. de Vries\inst{\ref{brusselsvrije}}
\and G. de Wasseige\inst{\ref{brusselsvrije}}
\and M. de With\inst{\ref{berlin}}
\and T. DeYoung\inst{\ref{michigan}}
\and J. C. D\'\i az-V\'elez\inst{\ref{madisonpac}}
\and H. Dujmovic\inst{\ref{skku}}
\and M. Dunkman\inst{\ref{pennphys}}
\and E. Dvorak\inst{\ref{southdakota}}
\and B. Eberhardt\inst{\ref{madisonpac}}
\and T. Ehrhardt\inst{\ref{mainz}}
\and P. Eller\inst{\ref{pennphys}}
\and P. A. Evenson\inst{\ref{bartol}}
\and S. Fahey\inst{\ref{madisonpac}}
\and A. R. Fazely\inst{\ref{southern}}
\and J. Felde\inst{\ref{maryland}}
\and K. Filimonov\inst{\ref{berkeley}}
\and C. Finley\inst{\ref{stockholmokc}}
\and A. Franckowiak\inst{\ref{zeuthen}}
\and E. Friedman\inst{\ref{maryland}}
\and A. Fritz\inst{\ref{mainz}}
\and T. K. Gaisser\inst{\ref{bartol}}
\and J. Gallagher\inst{\ref{madisonastro}}
\and E. Ganster\inst{\ref{aachen}}
\and S. Garrappa\inst{\ref{zeuthen}}
\and L. Gerhardt\inst{\ref{lbnl}}
\and K. Ghorbani\inst{\ref{madisonpac}}
\and T. Glauch\inst{\ref{munich}}
\and T. Gl\"usenkamp\inst{\ref{erlangen}}
\and A. Goldschmidt\inst{\ref{lbnl}}
\and J. G. Gonzalez\inst{\ref{bartol}}
\and D. Grant\inst{\ref{michigan}}
\and Z. Griffith\inst{\ref{madisonpac}}
\and M. G\"und\"uz\inst{\ref{bochum}}
\and C. Haack\inst{\ref{aachen}}
\and A. Hallgren\inst{\ref{uppsala}}
\and L. Halve\inst{\ref{aachen}}
\and F. Halzen\inst{\ref{madisonpac}}
\and K. Hanson\inst{\ref{madisonpac}}
\and D. Hebecker\inst{\ref{berlin}}
\and D. Heereman\inst{\ref{brusselslibre}}
\and K. Helbing\inst{\ref{wuppertal}}
\and R. Hellauer\inst{\ref{maryland}}
\and F. Henningsen\inst{\ref{munich}}
\and S. Hickford\inst{\ref{wuppertal}}
\and J. Hignight\inst{\ref{michigan}}
\and G. C. Hill\inst{\ref{adelaide}}
\and K. D. Hoffman\inst{\ref{maryland}}
\and R. Hoffmann\inst{\ref{wuppertal}}
\and T. Hoinka\inst{\ref{dortmund}}
\and B. Hokanson-Fasig\inst{\ref{madisonpac}}
\and K. Hoshina\inst{\ref{madisonpac},\ref{a}}
\and F. Huang\inst{\ref{pennphys}}
\and M. Huber\inst{\ref{munich}}
\and K. Hultqvist\inst{\ref{stockholmokc}}
\and M. H\"unnefeld\inst{\ref{dortmund}}
\and R. Hussain\inst{\ref{madisonpac}}
\and S. In\inst{\ref{skku}}
\and N. Iovine\inst{\ref{brusselslibre}}
\and A. Ishihara\inst{\ref{chiba}}
\and E. Jacobi\inst{\ref{zeuthen}}
\and G. S. Japaridze\inst{\ref{atlanta}}
\and M. Jeong\inst{\ref{skku}}
\and K. Jero\inst{\ref{madisonpac}}
\and B. J. P. Jones\inst{\ref{arlington}}
\and P. Kalaczynski\inst{\ref{aachen}}
\and W. Kang\inst{\ref{skku}}
\and A. Kappes\inst{\ref{munster}}
\and D. Kappesser\inst{\ref{mainz}}
\and T. Karg\inst{\ref{zeuthen}}
\and M. Karl\inst{\ref{munich}}
\and A. Karle\inst{\ref{madisonpac}}
\and U. Katz\inst{\ref{erlangen}}
\and M. Kauer\inst{\ref{madisonpac}}
\and A. Keivani\inst{\ref{pennphys}}
\and J. L. Kelley\inst{\ref{madisonpac}}
\and A. Kheirandish\inst{\ref{madisonpac}}
\and J. Kim\inst{\ref{skku}}
\and T. Kintscher\inst{\ref{zeuthen}}
\and J. Kiryluk\inst{\ref{stonybrook}}
\and T. Kittler\inst{\ref{erlangen}}
\and S. R. Klein\inst{\ref{lbnl},\ref{berkeley}}
\and R. Koirala\inst{\ref{bartol}}
\and H. Kolanoski\inst{\ref{berlin}}
\and L. K\"opke\inst{\ref{mainz}}
\and C. Kopper\inst{\ref{michigan}}
\and S. Kopper\inst{\ref{alabama}}
\and D. J. Koskinen\inst{\ref{copenhagen}}
\and M. Kowalski\inst{\ref{berlin},\ref{zeuthen}}
\and K. Krings\inst{\ref{munich}}
\and G. Kr\"uckl\inst{\ref{mainz}}
\and N. Kulacz\inst{\ref{edmonton}}
\and S. Kunwar\inst{\ref{zeuthen}}
\and N. Kurahashi\inst{\ref{drexel}}
\and A. Kyriacou\inst{\ref{adelaide}}
\and M. Labare\inst{\ref{gent}}
\and J. L. Lanfranchi\inst{\ref{pennphys}}
\and M. J. Larson\inst{\ref{maryland}}
\and F. Lauber\inst{\ref{wuppertal}}
\and J. P. Lazar\inst{\ref{madisonpac}}
\and K. Leonard\inst{\ref{madisonpac}}
\and M. Leuermann\inst{\ref{aachen}}
\and Q. R. Liu\inst{\ref{madisonpac}}
\and E. Lohfink\inst{\ref{mainz}}
\and C. J. Lozano Mariscal\inst{\ref{munster}}
\and L. Lu\inst{\ref{chiba}}
\and F. Lucarelli\inst{\ref{geneva}}
\and J. L\"unemann\inst{\ref{brusselsvrije}}
\and W. Luszczak\inst{\ref{madisonpac}}
\and J. Madsen\inst{\ref{riverfalls}}
\and G. Maggi\inst{\ref{brusselsvrije}}
\and K. B. M. Mahn\inst{\ref{michigan}}
\and Y. Makino\inst{\ref{chiba}}
\and K. Mallot\inst{\ref{madisonpac}}
\and S. Mancina\inst{\ref{madisonpac}}
\and I. C. Mari\c s\inst{\ref{brusselslibre}}
\and R. Maruyama\inst{\ref{yale}}
\and K. Mase\inst{\ref{chiba}}
\and R. Maunu\inst{\ref{maryland}}
\and K. Meagher\inst{\ref{madisonpac}}
\and M. Medici\inst{\ref{copenhagen}}
\and A. Medina\inst{\ref{ohio}}
\and M. Meier\inst{\ref{dortmund}}
\and S. Meighen-Berger\inst{\ref{munich}}
\and T. Menne\inst{\ref{dortmund}}
\and G. Merino\inst{\ref{madisonpac}}
\and T. Meures\inst{\ref{brusselslibre}}
\and S. Miarecki\inst{\ref{lbnl},\ref{berkeley}}
\and J. Micallef\inst{\ref{michigan}}
\and G. Moment\'e\inst{\ref{mainz}}
\and T. Montaruli\inst{\ref{geneva}}
\and R. W. Moore\inst{\ref{edmonton}}
\and M. Moulai\inst{\ref{mit}}
\and R. Nagai\inst{\ref{chiba}}
\and R. Nahnhauer\inst{\ref{zeuthen}}
\and P. Nakarmi\inst{\ref{alabama}}
\and U. Naumann\inst{\ref{wuppertal}}
\and G. Neer\inst{\ref{michigan}}
\and H. Niederhausen\inst{\ref{munich}}
\and S. C. Nowicki\inst{\ref{edmonton}}
\and D. R. Nygren\inst{\ref{lbnl}}
\and A. Obertacke Pollmann\inst{\ref{wuppertal}}
\and A. Olivas\inst{\ref{maryland}}
\and A. O'Murchadha\inst{\ref{brusselslibre}}
\and E. O'Sullivan\inst{\ref{stockholmokc}}
\and T. Palczewski\inst{\ref{lbnl},\ref{berkeley}}
\and H. Pandya\inst{\ref{bartol}}
\and D. V. Pankova\inst{\ref{pennphys}}
\and N. Park\inst{\ref{madisonpac}}
\and P. Peiffer\inst{\ref{mainz}}
\and C. P\'erez de los Heros\inst{\ref{uppsala}}
\and D. Pieloth\inst{\ref{dortmund}}
\and E. Pinat\inst{\ref{brusselslibre}}
\and A. Pizzuto\inst{\ref{madisonpac}}
\and M. Plum\inst{\ref{marquette}}
\and P. B. Price\inst{\ref{berkeley}}
\and G. T. Przybylski\inst{\ref{lbnl}}
\and C. Raab\inst{\ref{brusselslibre}}
\and A. Raissi\inst{\ref{christchurch}}
\and M. Rameez\inst{\ref{copenhagen}}
\and L. Rauch\inst{\ref{zeuthen}}
\and K. Rawlins\inst{\ref{anchorage}}
\and I. C. Rea\inst{\ref{munich}}
\and R. Reimann\inst{\ref{aachen}}
\and B. Relethford\inst{\ref{drexel}}
\and G. Renzi\inst{\ref{brusselslibre}}
\and E. Resconi\inst{\ref{munich}}
\and W. Rhode\inst{\ref{dortmund}}
\and M. Richman\inst{\ref{drexel}}
\and S. Robertson\inst{\ref{lbnl}}
\and M. Rongen\inst{\ref{aachen}}
\and C. Rott\inst{\ref{skku}}
\and T. Ruhe\inst{\ref{dortmund}}
\and D. Ryckbosch\inst{\ref{gent}}
\and D. Rysewyk\inst{\ref{michigan}}
\and I. Safa\inst{\ref{madisonpac}}
\and S. E. Sanchez Herrera\inst{\ref{edmonton}}
\and A. Sandrock\inst{\ref{dortmund}}
\and J. Sandroos\inst{\ref{mainz}}
\and M. Santander\inst{\ref{alabama}}
\and S. Sarkar\inst{\ref{oxford}}
\and S. Sarkar\inst{\ref{edmonton}}
\and K. Satalecka\inst{\ref{zeuthen}}
\and M. Schaufel\inst{\ref{aachen}}
\and P. Schlunder\inst{\ref{dortmund}}
\and T. Schmidt\inst{\ref{maryland}}
\and A. Schneider\inst{\ref{madisonpac}}
\and J. Schneider\inst{\ref{erlangen}}
\and L. Schumacher\inst{\ref{aachen}}
\and S. Sclafani\inst{\ref{drexel}}
\and D. Seckel\inst{\ref{bartol}}
\and S. Seunarine\inst{\ref{riverfalls}}
\and M. Silva\inst{\ref{madisonpac}}
\and R. Snihur\inst{\ref{madisonpac}}
\and J. Soedingrekso\inst{\ref{dortmund}}
\and D. Soldin\inst{\ref{bartol}}
\and M. Song\inst{\ref{maryland}}
\and G. M. Spiczak\inst{\ref{riverfalls}}
\and C. Spiering\inst{\ref{zeuthen}}
\and J. Stachurska\inst{\ref{zeuthen}}
\and M. Stamatikos\inst{\ref{ohio}}
\and T. Stanev\inst{\ref{bartol}}
\and A. Stasik\inst{\ref{zeuthen}}
\and R. Stein\inst{\ref{zeuthen}}
\and J. Stettner\inst{\ref{aachen}}
\and A. Steuer\inst{\ref{mainz}}
\and T. Stezelberger\inst{\ref{lbnl}}
\and R. G. Stokstad\inst{\ref{lbnl}}
\and A. St\"o\ss l\inst{\ref{chiba}}
\and N. L. Strotjohann\inst{\ref{zeuthen}}
\and T. Stuttard\inst{\ref{copenhagen}}
\and G. W. Sullivan\inst{\ref{maryland}}
\and M. Sutherland\inst{\ref{ohio}}
\and I. Taboada\inst{\ref{georgia}}
\and F. Tenholt\inst{\ref{bochum}}
\and S. Ter-Antonyan\inst{\ref{southern}}
\and A. Terliuk\inst{\ref{zeuthen}}
\and S. Tilav\inst{\ref{bartol}}
\and L. Tomankova\inst{\ref{bochum}}
\and C. T\"onnis\inst{\ref{skku}}
\and S. Toscano\inst{\ref{brusselsvrije}}
\and D. Tosi\inst{\ref{madisonpac}}
\and M. Tselengidou\inst{\ref{erlangen}}
\and C. F. Tung\inst{\ref{georgia}}
\and A. Turcati\inst{\ref{munich}}
\and R. Turcotte\inst{\ref{aachen}}
\and C. F. Turley\inst{\ref{pennphys}}
\and B. Ty\inst{\ref{madisonpac}}
\and E. Unger\inst{\ref{uppsala}}
\and M. A. Unland Elorrieta\inst{\ref{munster}}
\and M. Usner\inst{\ref{zeuthen}}
\and J. Vandenbroucke\inst{\ref{madisonpac}}
\and W. Van Driessche\inst{\ref{gent}}
\and D. van Eijk\inst{\ref{madisonpac}}
\and N. van Eijndhoven\inst{\ref{brusselsvrije}}
\and S. Vanheule\inst{\ref{gent}}
\and J. van Santen\inst{\ref{zeuthen}}
\and M. Vraeghe\inst{\ref{gent}}
\and C. Walck\inst{\ref{stockholmokc}}
\and A. Wallace\inst{\ref{adelaide}}
\and M. Wallraff\inst{\ref{aachen}}
\and N. Wandkowsky\inst{\ref{madisonpac}}
\and T. B. Watson\inst{\ref{arlington}}
\and C. Weaver\inst{\ref{edmonton}}
\and M. J. Weiss\inst{\ref{pennphys}}
\and J. Weldert\inst{\ref{mainz}}
\and C. Wendt\inst{\ref{madisonpac}}
\and J. Werthebach\inst{\ref{madisonpac}}
\and S. Westerhoff\inst{\ref{madisonpac}}
\and B. J. Whelan\inst{\ref{adelaide}}
\and N. Whitehorn\inst{\ref{ucla}}
\and K. Wiebe\inst{\ref{mainz}}
\and C. H. Wiebusch\inst{\ref{aachen}}
\and L. Wille\inst{\ref{madisonpac}}
\and D. R. Williams\inst{\ref{alabama}}
\and L. Wills\inst{\ref{drexel}}
\and M. Wolf\inst{\ref{munich}}
\and J. Wood\inst{\ref{madisonpac}}
\and T. R. Wood\inst{\ref{edmonton}}
\and K. Woschnagg\inst{\ref{berkeley}}
\and G. Wrede\inst{\ref{erlangen}}
\and D. L. Xu\inst{\ref{madisonpac}}
\and X. W. Xu\inst{\ref{southern}}
\and Y. Xu\inst{\ref{stonybrook}}
\and J. P. Yanez\inst{\ref{edmonton}}
\and G. Yodh\inst{\ref{irvine}}
\and S. Yoshida\inst{\ref{chiba}}
\and T. Yuan\inst{\ref{madisonpac}}\\
\vspace{0.15cm}
J.~Nordin\inst{\ref{berlin}}
}}
\institute{
{\tiny
Astrophysics Research Centre, School of Mathematics and Physics, Queen's University Belfast, Belfast, BT7 1NN,  UK\label{Belfast} 
\and Tuorla Observatory, Department of Physics and Astronomy, University of Turku, FI-20014 Turku, Finland\label{Turku}
\and Institute of Astronomy, University of Hawaii, 2680 Woodlawn Drive, Honolulu, Hawaii 96822, USA\label{Honolulu} 
\and  Max-Planck-Institut fur Extraterrestrische Physik, Giessenbachstrasse 1, D-85748, Garching, Germany\label{MPE} 
\and Minnesota Institute for Astrophysics, University of Minnesota, Minneapolis, MN 55454, USA\label{minnesota} 
\and III. Physikalisches Institut, RWTH Aachen University, D-52056 Aachen, Germany \label{aachen} 
\and Department of Physics, University of Adelaide, Adelaide, 5005, Australia \label{adelaide} 
\and Dept. of Physics and Astronomy, University of Alaska Anchorage, 3211 Providence Dr., Anchorage, AK 99508, USA \label{anchorage} 
\and Dept. of Physics, University of Texas at Arlington, 502 Yates St., Science Hall Rm 108, Box 19059, Arlington, TX 76019, USA \label{arlington} 
\and CTSPS, Clark-Atlanta University, Atlanta, GA 30314, USA \label{atlanta} 
\and School of Physics and Center for Relativistic Astrophysics, Georgia Institute of Technology, Atlanta, GA 30332, USA \label{georgia} 
\and Dept. of Physics, Southern University, Baton Rouge, LA 70813, USA \label{southern} 
\and Dept. of Physics, University of California, Berkeley, CA 94720, USA \label{berkeley} 
\and Lawrence Berkeley National Laboratory, Berkeley, CA 94720, USA \label{lbnl} 
\and Institut f\"ur Physik, Humboldt-Universit\"at zu Berlin, D-12489 Berlin, Germany \label{berlin} 
\and Fakult\"at f\"ur Physik \& Astronomie, Ruhr-Universit\"at Bochum, D-44780 Bochum, Germany \label{bochum} 
\and Universit\'e Libre de Bruxelles, Science Faculty CP230, B-1050 Brussels, Belgium \label{brusselslibre} 
\and Vrije Universiteit Brussel (VUB), Dienst ELEM, B-1050 Brussels, Belgium \label{brusselsvrije} 
\and Dept. of Physics, Massachusetts Institute of Technology, Cambridge, MA 02139, USA \label{mit} 
\and Dept. of Physics and Institute for Global Prominent Research, Chiba University, Chiba 263-8522, Japan \label{chiba} 
\and Dept. of Physics and Astronomy, University of Canterbury, Private Bag 4800, Christchurch, New Zealand \label{christchurch} 
\and Dept. of Physics, University of Maryland, College Park, MD 20742, USA \label{maryland} 
\and Dept. of Astronomy, Ohio State University, Columbus, OH 43210, USA \label{ohioastro} 
\and Dept. of Physics and Center for Cosmology and Astro-Particle Physics, Ohio State University, Columbus, OH 43210, USA \label{ohio} 
\and Niels Bohr Institute, University of Copenhagen, DK-2100 Copenhagen, Denmark \label{copenhagen} 
\and Dept. of Physics, TU Dortmund University, D-44221 Dortmund, Germany \label{dortmund} 
\and Dept. of Physics and Astronomy, Michigan State University, East Lansing, MI 48824, USA \label{michigan} 
\and Dept. of Physics, University of Alberta, Edmonton, Alberta, Canada T6G 2E1 \label{edmonton} 
\and Erlangen Centre for Astroparticle Physics, Friedrich-Alexander-Universit\"at Erlangen-N\"urnberg, D-91058 Erlangen, Germany \label{erlangen} 
\and Physik-department, Technische Universit\"at M\"unchen, D-85748 Garching, Germany \label{munich} 
\and D\'epartement de physique nucl\'eaire et corpusculaire, Universit\'e de Gen\`eve, CH-1211 Gen\`eve, Switzerland \label{geneva} 
\and Dept. of Physics and Astronomy, University of Gent, B-9000 Gent, Belgium \label{gent} 
\and Dept. of Physics and Astronomy, University of California, Irvine, CA 92697, USA \label{irvine} 
\and Dept. of Physics and Astronomy, University of Kansas, Lawrence, KS 66045, USA \label{kansas} 
\and SNOLAB, 1039 Regional Road 24, Creighton Mine 9, Lively, ON, Canada P3Y 1N2 \label{snolab} 
\and Department of Physics and Astronomy, UCLA, Los Angeles, CA 90095, USA \label{ucla} 
\and Dept. of Astronomy, University of Wisconsin, Madison, WI 53706, USA \label{madisonastro} 
\and Dept. of Physics and Wisconsin IceCube Particle Astrophysics Center, University of Wisconsin, Madison, WI 53706, USA \label{madisonpac} 
\and Institute of Physics, University of Mainz, Staudinger Weg 7, D-55099 Mainz, Germany \label{mainz} 
\and Department of Physics, Marquette University, Milwaukee, WI, 53201, USA \label{marquette} 
\and Institut f\"ur Kernphysik, Westf\"alische Wilhelms-Universit\"at M\"unster, D-48149 M\"unster, Germany \label{munster} 
\and Bartol Research Institute and Dept. of Physics and Astronomy, University of Delaware, Newark, DE 19716, USA \label{bartol} 
\and Dept. of Physics, Yale University, New Haven, CT 06520, USA \label{yale} 
\and Dept. of Physics, University of Oxford, Parks Road, Oxford OX1 3PQ, UK \label{oxford} 
\and Dept. of Physics, Drexel University, 3141 Chestnut Street, Philadelphia, PA 19104, USA \label{drexel} 
\clearpage
\and Physics Department, South Dakota School of Mines and Technology, Rapid City, SD 57701, USA \label{southdakota} 
\and Dept. of Physics, University of Wisconsin, River Falls, WI 54022, USA \label{riverfalls} 
\and Dept. of Physics and Astronomy, University of Rochester, Rochester, NY 14627, USA \label{rochester} 
\and Oskar Klein Centre and Dept. of Physics, Stockholm University, SE-10691 Stockholm, Sweden \label{stockholmokc} 
\and Dept. of Physics and Astronomy, Stony Brook University, Stony Brook, NY 11794-3800, USA \label{stonybrook} 
\and Dept. of Physics, Sungkyunkwan University, Suwon 16419, Korea \label{skku} 
\and Dept. of Physics and Astronomy, University of Alabama, Tuscaloosa, AL 35487, USA \label{alabama} 
\and Dept. of Astronomy and Astrophysics, Pennsylvania State University, University Park, PA 16802, USA \label{pennastro} 
\and Dept. of Physics, Pennsylvania State University, University Park, PA 16802, USA \label{pennphys} 
\and Dept. of Physics and Astronomy, Uppsala University, Box 516, S-75120 Uppsala, Sweden \label{uppsala} 
\and Dept. of Physics, University of Wuppertal, D-42119 Wuppertal, Germany \label{wuppertal} 
\and DESY, D-15738 Zeuthen, Germany \label{zeuthen} 
\and Earthquake Research Institute, University of Tokyo, Bunkyo, Tokyo 113-0032, Japan\label{a}
}}

   \titlerunning{IceCube Neutrinos with PS1}
   \authorrunning{Kankare et al.}

   \maketitle


\twocolumn[\begin{center}
{\small\sffamily\bfseries\MakeUppercase ABSTRACT}
\end{center}
{\small
In order to identify the sources of the observed diffuse high-energy neutrino flux, it is crucial to discover their electromagnetic counterparts. To increase the sensitivity of detecting counterparts of transient or variable sources by telescopes with a limited field of view, IceCube began releasing alerts for single high-energy ($E_{\nu}>60$\,TeV) neutrino detections with sky localisation regions of order 1\,degree radius in 2016. We used Pan-STARRS1 to follow-up five of these alerts during 2016$-$2017 to search for any optical transients that may be related to the neutrinos.  Typically 10$-$20 faint ($m_{i_{\rm P1}} \lesssim 22.5$ mag) extragalactic transients are found within the Pan-STARRS1 footprints and are generally consistent with being unrelated field supernovae (SNe) and AGN. We looked for unusual properties of the detected transients, such as temporal coincidence of explosion epoch with the IceCube timestamp, or other peculiar light curve and physical properties. We found only one transient that had properties worthy of a specific follow-up. In the Pan-STARRS1 imaging for IceCube-160427A (probability to be of astrophysical origin of $\sim 50$\%), we found a SN PS16cgx, located at 10.0\arcmin\ from the nominal IceCube direction. Spectroscopic observations of PS16cgx showed that it was an H-poor SN at redshift $z = 0.2895 \pm 0.0001$. The spectra and light curve resemble some high-energy Type Ic SNe, raising the possibility of a jet driven SN with an explosion epoch temporally coincident with the neutrino detection. However, distinguishing Type Ia and Type Ic SNe at this redshift is notoriously difficult. Based on all available data we conclude that the transient is more likely to be a Type Ia with relatively weak \SiII\ absorption and a fairly normal rest-frame $r-$band light curve. If, as predicted, there is no high-energy neutrino emission from Type Ia SNe, then PS16cgx must be a random coincidence, and unrelated to the IceCube-160427A. We find no other plausible optical transient for any of the five IceCube events observed down to a 5$\sigma$ limiting magnitude of $m_{i_{\rm P1}} \approx 22$ mag, between 1 day and 25 days after detection.}
\vspace{5mm}\\
{\small\sffamily\bfseries\MakeUppercase Key words.}
{\small astroparticle physics -- neutrinos -- supernovae: general -- supernovae: individual (PS16cgx)}\\
\vspace{15mm}
]

\section{Introduction}

High-energy neutrinos are produced in hadronic interactions in the most violent sources in the Universe and hence provide a smoking-gun signature for the origin of cosmic rays. Furthermore, neutrinos are capable of escaping even the densest environments such as stellar cores and black hole accretion discs. Along with gravitational waves \citep{abbott16,abbott17}, they  carry information of these environments that neither photons or charged particles can provide. After the first detection of a diffuse flux of high-energy astrophysical neutrinos \citep{aartsen13}, the most pressing question in the field of neutrino astronomy is the astrophysical origin of the particles. No significant cluster in space or time has been found in un-triggered searches of neutrino data yet \citep{aartsen17a} and the isotropic distribution of the neutrinos points to an extragalactic origin. Multi-wavelength observations are the key to identify neutrino sources through the detection of an electromagnetic counterpart. Recently, the high-energy 290 TeV neutrino event IceCube-170922A has been associated with a flare in the blazar TXS 0506+056 at 3$\sigma$ significance \citep{aartsen18a,aartsen18b}. Photohadronic processes in the cores of active galaxies have previously been proposed as the sources of these high-energy neutrinos \citep{padovani14,petropoulou15} and the IceCube-170922A association is an intriguing development in the field. The only other identified extragalactic source of neutrinos is the core-collapse supernova (CCSN) 1987A. Undisputed detection of 20 neutrinos were made by two experiments in the energy range of 7.5 to 40 MeV \citep{hirata87,bionta87}, corresponding to the thermal regime in the collapsing core of the massive star that produced SN 1987A. 

Gamma-ray bursts (GRB) have also been suggested as source candidates for the highest-energy cosmic rays and high-energy neutrinos \citep{waxman95,waxman97}. \citet{aartsen16} suggest that the bulk of the known population of bright GRBs can be excluded as the main contributor to IceCube's diffuse neutrino flux. However, a large population of low-luminosity GRBs might contribute significantly to the observed flux~\citep{murase06,murase13}. While highly-relativistic jets can explain gamma-ray bright bursts, `choked' jets within the star may explain relativistic SNe and low-luminosity GRBs, giving a unified picture of GRBs and GRB-SNe \citep{meszaros01,senno16}. In choked-jet SNe, the jet fails to penetrate the surface of the progenitor star. Neutrinos and gamma rays can be produced by hadronic processes in the jet. While neutrinos can leave the dense environment, gamma rays are absorbed. This scenario could be physically probed by the detection of high-energy neutrinos in coincidence with SNe that could harbour relativistic jets~\citep{kowalski07,esmaili18}. Such  neutrino emission is expected in a relatively short time window ($\sim$100 s) after core collapse. The signature of a choked-jet SN could therefore be a high-energy neutrino signal followed by a young CCSN. Explosions of CCSNe originate from stars more massive than $M \approx 8 M_{\odot}$ and are classified based on their spectroscopic features \citep{filippenko97}. Type II SNe are H-rich and Type Ib/c SNe H-poor. Furthermore, Type Ib SNe show He in their spectra whereas Type Ic SNe lack these features. Type Ic SNe are thought to originate from massive stars that have lost their outer envelopes \citep[Wolf-Rayet stars, e.g.][]{crowther07} or less massive stars in binary systems \citep[e.g.][]{podsiadlowski92}. High-energy neutrino production in Type Ic SNe has been postulated to be possible, as a result of diffusive shock acceleration of protons in relativistic jets powered by a central engine \citep{razzaque05,ando05,tamborra16,senno16}. Those protons can produce pions in interactions with ambient photon or matter fields. Charged pions produce neutrinos in their decay chain. The mechanism follows the GRB-fireball phenomenology \citep{piran99}, but unlike conventional long GRBs with $\Gamma$ factors of $10^2-10^3$, the culprit SNe are assumed to harbour fewer relativistic jets. Such jets can be either choked or emerge, thereby producing low-luminosity or long-duration GRBs (potentially undetected by current facilities). Furthermore, it is expected that Type IIn SNe may also produce high-energy neutrinos arising from the shock interaction of the expanding SN ejecta and circumstellar medium. A Type IIn SN PTF12csy was discovered within the confidence region of two high-energy neutrinos detected by IceCube on 2012 March 30. However, the Pan-STARRS1 data showed the SN to be at least 158 days old and therefore likely unrelated \citep{aartsen15}. Contrary to CCSNe, Type Ia SNe are thermonuclear explosions of white dwarf stars and are not expected to produce high-energy neutrinos.

In the following sections, we briefly introduce the IceCube realtime neutrino search in Section~\ref{sect:NeutrinoSearch} followed by a detailed discussion of the neutrino event IceCube-160427A in Section~\ref{sect:NeutrinoEvent}. Section~\ref{sect:Supernova} presents the observations and interpretation of a young SN PS16cgx discovered in the confidence region of the neutrino event. Summary of Pan-STARRS1 observations of other IceCube alerts is provided in Section~\ref{sect:summ}. The probability to associate neutrino emitters with Pan-STARRS1 is discussed in Section~\ref{sect:Discussion} and the results of the paper are summarised in Section~\ref{sect:Conclusion}. In all cases we adopt a calibration to standard cosmology with $H_{0} = 70$ km s$^{-1}$ Mpc$^{-1}$, $\Omega_{\mathrm{m}} = 0.3$, and $\Omega_{\Lambda} = 0.7$. 

\section{The IceCube realtime neutrino search}
\label{sect:NeutrinoSearch}

The IceCube neutrino Observatory is a 1 km$^3$ water Cherenkov detector located in the clear glacial ice at the geographic South Pole. 5160 digital optical modules (DOMs) detect light emission in the ice at a depth ranging from 1450 to 2450 m~\citep{abbasi09,aartsen13b,aartsen17c}. Charged-current interactions of muon neutrinos in the ice produce muons, which are detected using emitted Cherenkov light, and which -- at TeV energies -- travel long distances of several kilometres in the Antarctic ice \citep{chirkin04}. The resulting long lever arm of those `track events' gives a good reconstruction performance with median angular resolution of $\Delta \Psi \lesssim 1^\circ$. Charged-current interactions of electron or tau neutrinos, as well as neutral-current interactions of any neutrino type, produce `shower events' with almost spherically symmetric light emission resulting in an angular resolution of only about 10$^\circ$.

The IceCube realtime programme selects neutrinos with a high probability of being of astrophysical origin, reconstructs their direction and notifies ground and space-based observatories with a median latency of 33\,s to perform multi-wavelengths follow-up observations aiming for the detection of an electro-magnetic counterpart \citep[see][for a detailed description]{aartsen17b}. 

Starting in late 2008, IceCube implemented an optical follow-up programme on multiplets. In order to suppress the background from atmospheric neutrinos, a multiplet of at least two neutrinos within 100 s and angular separation of 3\fdg5 or less was required to trigger an alert \citep{abbasi12,aartsen17b}. A number of alerts were sent to the Palomar Transient Factory (PTF) and \textit{Swift} \citep{evans15} and which has not resulted in a compelling coincidence with an electromagnetic source \citep{aartsen15}. In 2016, IceCube began releasing single neutrinos in two neutrino candidate streams \citep{aartsen17b} called High-Energy Starting Events (HESE) and Extremely High-energy Events (EHE), which are the focus of this paper. 

 The HESE sample consists of events which have their interaction vertices inside the detector volume and are selected by using the outer layers of the detector as a veto to suppress the background of atmospheric muons, which are produced in cosmic-ray air showers and enter the detector from above~\citep{aartsen14}. The veto is only efficient at high energies and is therefore combined with a requirement on the charge deposited in the detector of $Q > 6000$ photoelectrons, which roughly corresponds to an energy deposited in the detector of 60 TeV. On average, four HESE tracks are detected per year, of which one is expected to be of astrophysical origin. Since April 2016, these have been made public in realtime using the Astrophysical Multimessenger Observatory Network \citep[AMON;][]{smith13,aartsen17b} 
 and via the Gamma-Ray Coordinates Network\footnote{\url{https://gcn.gsfc.nasa.gov/}, \url{https://gcn.gsfc.nasa.gov/amon\_hese\_events.html} and \url{https://gcn.gsfc.nasa.gov/amon\_ehe\_events.html}} (GCN).

EHE events, which do not have to start inside the detector but can be through-going, are selected by a combination of energy and zenith angle \citep[see][for a detailed description]{aartsen17b}. These neutrinos typically have energies $>200$\,TeV. The public EHE stream was activated in July 2016 and is expected to produce four alerts per year with an average signal purity of 50\%. In this paper we introduce the neutrino detection electromagnetic counterpart search with Pan-STARRS1 with a focus on the first publicly released event IceCube-160427A.

\section{IceCube-160427A}
\label{sect:NeutrinoEvent}

The first HESE event after the start of the programme was detected on 2016 April 27 at 05:52:32 UT (MJD 57505.24481) and released to the public via a GCN notice\footnote{\url{https://gcn.gsfc.nasa.gov/notices_amon/67093193_127853.amon}}. A summary was published as a GCN circular a few days later \citep{blaufuss16}.

Several instruments followed-up the IceCube event searching for a counterpart in various wavelengths. In gamma rays, observations were performed by the First G-APD Cherenkov Telescope \citep[FACT;][]{biland16}, the Interplanetary Network \citep[IPN\footnote{\url{www.ssl.berkeley.edu/ipn3/}};][]{svinkin16}, the Very Energetic Radiation Imaging Telescope Array System \citep[VERITAS;][]{mukherjee16}, the High-Altitude Water Cherenkov Gamma-Ray Observatory \citep[HAWC;][]{taboada16}, the Fermi Large Area Telescope \citep[Fermi-LAT;][]{vianello16} and the Fermi Gamma-Ray Burst Monitor \citep[Fermi-GBM;][]{blackburn16}, and in optical wavelength by the Intermediate Palomar Transient Factory \citep[iPTF;][]{singer16a}, the Panoramic Survey Telescope and Rapid Response System 1 \citep[Pan-STARRS1;][]{smartt16c} and the Mobile Astronomical System of Telescope-Robots \citep[MASTER;][]{lipunov16a}. Except a young SN discovered by Pan-STARRS1, no compelling counterpart detection was reported.

\subsection{Direction}

An angular reconstruction is applied to the HESE events at the South Pole, which is used to send a first notification. More time consuming reconstructions are applied, once the event data arrive in the north and provide a more accurate estimate on the neutrino direction within a few hours. A full likelihood scan is applied on a narrow grid with about 0\fdg06 distance between the grid points. The resulting map of the likelihood landscape allows us to find the global minimum and the error contours at a given confidence level. The calculation of the updated direction was not fully automated yet when the first public HESE track was detected, which led to an additional delay. Table~\ref{tab:alert} shows the initially published direction 81 s after the detection of the neutrino compared to the updated position 17.5 h later. The position moved by 2\fdg6.

\begin{figure}
\centering
\includegraphics[width=\columnwidth]{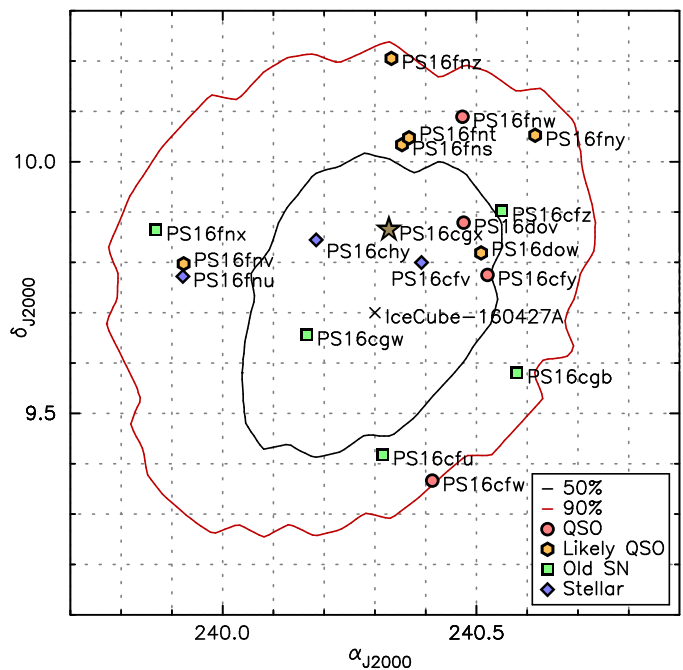}
\caption{The black (red) line shows the 50\% (90\%) confidence region. The confidence region likelihood ratios were determined using Monte Carlo re-simulations of the event, taking into account uncertainties in the optical properties of the deep glacial ice. The best fit position is shown as a $\times$ mark and the position of PS16cgx is indicated with a star symbol. Furthermore, the other faint transients discovered within the final 90\% probability containment by Pan-STARRS1 are also shown and contextually classified based on their host or precursor information as labeled in the legend.}
\label{fig:llh}
\end{figure}

The likelihood scan was repeated after cleaning out late photomultiplier hits from afterpulsing, an effect not currently described by the track likelihood hypothesis. This improves the likelihood description and leads to a better fit. The final best fit position and the 50 and 90\% confidence regions are shown in Fig.~\ref{fig:llh}. The positions of the contour lines were determined using re-simulations of similar events while also varying the allowed models of the optical properties of the deep glacial ice in order to obtain a conservative range. The minimum yields our best estimate of the event direction: $\alpha_{\mathrm{J2000}} = 240\fdg3$ and $\delta_{\mathrm{J2000}} = 9\fdg7$, with a 90\% error ellipse contour with semi major (minor) axis of 0\fdg54 (0\fdg45). 

\begin{table}
\setlength{\tabcolsep}{5pt}
\caption{Coordinate revisions of IceCube-160427A.}
\begin{tabular}{cccccc}
\hline
\hline
Time$^{a}$  & $\alpha_{\mathrm{J2000}}$ & $\delta_{\mathrm{J2000}}$ & 50\% error$^{b}$ & 90\% error & $\Delta\Psi_{\textrm{16cgx}}^{c}$ \\
 & ($^{\circ}$) & ($^{\circ}$) & ($^{\circ}$) & ($^{\circ}$) & ($^{\circ}$) \\
\hline
81 s   & 239.7 &  6.9 & 1.6 & 8.9 & 3.03 \\
17.5 h & 240.6 &  9.3 & $-$ & 0.6 & 0.63 \\
Final & 240.3 & 9.7 & 0.22$-$0.34 & 0.45$-$0.54 & 0.17 \\
\hline
\label{tab:alert}
\end{tabular}
\begin{flushleft}
$^{a}$ Time relative to the neutrino detection.\\
$^{b}$ The range for the error radius in the last row gives the range between major and minor semi-axis when fitting the likelihood landscape with an ellipse.\\
$^{c}$ Distance of the neutrino position to PS16cgx.
\end{flushleft}
\end{table}

\subsection{Energy}

The total charge deposited in the detector amounts to $1.9 \times 10^{4}$ photoelectrons, corresponding to a deposited energy of 134 TeV. Using Monte Carlo simulations, we estimate a lower bound on the true energy of the primary neutrino. 
In the following we assume a spectral index of $-2.58$ for the astrophysical flux \citep{kopper15}, which is the best fit value obtained from the 4 yr HESE sample. We then simulate events with the same starting position in the detector (within $\pm$30 m horizontally and $\pm$15 m vertically), direction (within $2^\circ$) and deposited energy (within $\pm$20\%). By comparison of IceCube-160427A to the simulated events we estimate the true neutrino energy to be larger than 130 TeV (230 TeV) at 90\% (50\%) confidence (which, as expected, is similar to the deposited energy of the event). The result is only weakly dependent on the chosen spectral index \citep[see][]{aartsen18b}.

\subsection{Signal purity}

On average 25\% of the events in the HESE realtime stream are expected to be astrophysical neutrinos. The HESE events are characterised by the signal-trackness parameter, which is a number between 0 and 1, characterising how likely the event is both astrophysical as well as track-like, opposed to shower-like \citep[see][for more details]{aartsen17b}. For IceCube-160427A, the signal-trackness parameter is $0.92$. We estimate the fraction of astrophysical signal events in the alert sample for events with a similar or larger deposited charge and which enter the detector from a similar zenith direction. We assume a power-law signal spectrum. We find a fraction of 46$-$58\% depending on the assumed spectral index ranging from $\gamma=-2.19$, found in the analysis of the diffuse muon neutrino flux~\citep{haack17}, to $\gamma=-2.92$, obtained in an analysis of high-energy starting-tracks~\citep{kopper17}. The fraction is similar to that of IceCube-170922A, which had a higher reconstructed energy but passed the EHE instead of the HESE stream selection criteria. The HESE event selection uses the outer detector layers as veto to select track events starting inside the instrumented detector volume.

\subsection{Offline neutrino search}

We investigate whether the HESE event was accompanied by an excess of low-energy neutrino events. We applied a maximum likelihood analysis following the procedure in \citet{meagher17} to search for neutrino emission from a point source in the direction of PS16cgx on various time scales of 10, 100, 1000, 10,000 and 100,000\,s centred on the time of the HESE event. The chosen time windows cover the shock propagation and shock breakout times scales estimated by~\citet{kistler13}. The HESE event was removed from the likelihood analysis to avoid a bias in the p-value calculation. No significant excess was found in any of the tested time windows leading to a 90\% energy flux upper limit of $3.7\times 10^{-5}$ TeV cm$^{-2}$ over the 100,000 s time window in an approximate energy range of 1 TeV to 1 PeV assuming a neutrino spectrum following an $E^{-2}$ dependence. The non-detection of additional neutrinos is not surprising, since we expect sources of two detectable neutrinos at a median redshift of 0.06 \citep[see Figure 8 of][]{aartsen17d} assuming the sources follow redshift evolution of the star-formation rate.

\section{Pan-STARRS1 observations of IceCube EHE and HESE alerts}
\label{sect:Supernova}

The 1.8-m Pan-STARRS1 telescope \citep{chambers16} is located at Haleakala Observatory, Hawaii, USA. It is equipped with a  1.4 gigapixel mosaic CCD camera, called Gigapixel Camera 1 (GPC1). The field of view of GPC1 is $\sim$7 deg$^{2}$ and has a pixel size of 0.25\arcsec/pix. The Pan-STARRS1 data are processed automatically by the real-time Pan-STARRS1 Image Processing Pipeline \citep[IPP;][]{magnier16a}. The photometric calibration of the Pan-STARRS1 images is described in \citet{magnier16b}. Each image has a reference sky subtracted from it to produce a calibrated difference image. After this reduction, calibration and image subtraction in Hawaii, the catalogues of all detections are transferred to Queen's University Belfast where they are ingested into a MySQL database, assimilated into objects, and undergo a series of filters to produce a clean stream of real astrophysical transients \citep{wright15}. They are cross-matched with a large custom-built database of star and galaxy catalogues and annotated with associations to known sources. The processing is described in more detail in \citet{smartt16a} for searches for optical signatures of gravitational wave sources. Similar techniques were employed here. The diameter of the field of view of Pan-STARRS1 is 3$^{\circ}$ \citep{chambers16} and hence one pointing typically encloses the full 90\% confidence area estimated by IceCube, which tend to have radii of around $\sim$ 0\fdg5$-$1$^{\circ}$. We have typically employed a strategy of multiple dithered exposures in \ips\ or \rps\ filters as were done in the Pan-STARRS1 Science Consortium Medium Deep Fields \citep[for details see][]{rest14,mccrum15,lunnan18,scolnic18}. These dithered exposures are usually $8\times240$ s long. They are stacked together to make one 1920 s exposure although we also process the individual 8 exposures. We carried out a concerted campaign to observe the error boxes of EHE and HESE events in 2016 and 2017 with Pan-STARRS1, and observed five in total (IceCube-160427A, IceCube-160731A, IceCube-161103A, IceCube-170321A, and IceCube-171106A). We focus here on details of the first event released (IceCube-160427A) since that was the only event in which we found a young, rising extragalactic transient with a possible temporal coincidence with the neutrino detection time. The results from the rest of the events are summarised in Section~\ref{sect:summ}. 

The Pan-STARRS1 telescope began observing the field of IceCube-160427A in the \ips-band on MJD 57508.5 (2016 April 30.5 UT), 3.3 d after the IceCube detection, and the observations were repeated in the \ips-band on 4 subsequent nights. Individual dithered images were stacked together and the 3$\pi$ Steradian Survey reference sky  \citep[see][]{chambers16} was subtracted from these nightly stacked images. 
The 3$\pi$ reference sky is constructed from the median combination of multiple individual exposures that are of length 43 (in filter \gps), 40 (\rps), 45 (\ips), 40 (\zps) and 30 (\yps) s. Typically there are around 12 individual exposures in each filter at each point in the sky that are combined to form the reference sky, which were taken over the period between 2009$-$2014. However, the number of exposures per point on the sky is a distribution covering the range between 5 and 25 \citep[see Fig. 16 of][]{chambers16}. For follow-up of IceCube neutrino alerts, the \ips\ filter was typically used and, in most cases, the number of exposures in the \ips\ reference sky is around 16. The typical depth of the reference sky is $m_{i_{\rm{P1}}} \approx 23.1$ mag \citep[see Table 11 and Fig. 17 of][]{chambers16}.

After subtraction, the stacked images typically had a 5$\sigma$ point-source image depth of $m_{i_{\rm{P1}}} \approx 22.5$ mag. The initial results, based on the first revision, of the IceCube-160427A coordinates were reported by \citet{smartt16c}. In this paper, we revise these results with the final IceCube-160427A position. As an important null result, we report that no persistent bright ($m_{i} \lesssim$ 20 mag) transients were found in the IceCube-160427A field over a period of +3.3 to 45 days. While we did not cover the first 3 days, we rule out any relatively bright SN-like transient that exploded at a time coincident with IceCube-160427A. Other optical surveys by our team and others have found a similar null result for IceCube events, in that there are no bright candidates with SN-like light curves in the field \citep[e.g.][]{lipunov16a,lipunov16b,singer16a,singer16b}. 

For IceCube-160427A, we found 19 faint optical transients  within the 90\% probability containment of the IceCube-160427A position: 4 variable known quasi-stellar objects (QSOs), that is, extremely luminous active galactic nuclei (AGNs), based on the Sloan Digital Sky Survey (SDSS) Data Release 12 \citep[DR12;][]{alam15} archival spectra; 6 likely QSOs based on the host colour; 6 SN candidates; and 3 faint sources that are likely either faint Galactic variable stars or QSO variability, see Table~\ref{tab:ps1}. The SN candidates are transients that are either offset from or coincident with an apparent host galaxy which is not a known QSO or AGN. Five of the SN candidates show flat or declining light curves, consistent with SNe at peak or post maximum. However, one of the candidates, PS16cgx, shows a rising light curve of 0.4 mag over 2 days suggesting that it is a young SN and therefore possibly having an explosion epoch that is temporally consistent with IceCube-160427A. PS16cgx was located at $\alpha_{\mathrm{J2000}}$ = 240\fdg32755 and $\delta_{\mathrm{J2000}}$ = +9\fdg86478 (16$^{\mathrm{h}}$01$^{\mathrm{m}}$18\fs61 +09$^{\circ}$51\arcmin53\farcs1), 10.0\arcmin\ from the best-fit IceCube-160427A direction and consistent with the neutrino position within the uncertainties at 50\% confidence.

\begin{table*}
\caption{19 transients discovered within the final 90\% probability containment of the IceCube-160427A neutrino detection. The time of detection of IceCube-160427A was MJD=57505.24482.}
\begin{tabular}{lcccccccc}
\hline
\hline
\multicolumn{1}{c}{Transient} & $\alpha_{\mathrm{J2000}}$ & $\delta_{\mathrm{J2000}}$ & Disc. epoch & $m_{\mathrm{disc},i}$ & Type & $z^{a}$ & $m_{\mathrm{host},i}$ &  $\Delta\Psi^{b}$\\
 & (hh:mm:ss.ss) & (dd:mm:ss.s) & (MJD) & (mag) & & & (mag) & ($^{\circ}$) \\
\hline
PS16cfv & 16:01:34.00 & +09:47:59.1 & 57508.49664 & 21.76 & Stellar & $-$ & 21.21 & 0.13 \\
PS16cgw & 16:00:39.69 & +09:39:21.2 & 57508.49664 & 21.80 & Old SN & 0.5792$\pm$0.1889 & 20.58 & 0.14 \\
PS16cgx & 16:01:18.60 & +09:51:53.1 & 57508.49664 & 21.84 & Young SN & 0.2895$\pm$0.0001 & 19.40 & 0.17 \\
PS16chy & 16:00:44.22 & +09:50:40.8 & 57508.49664 & 21.73 & Stellar & $-$ & 20.45 & 0.18 \\
PS16cfy & 16:02:05.21 & +09:46:30.4 & 57508.49664 & 21.67 & QSO & 2.3438$\pm$0.0004 & 21.20 & 0.23 \\
PS16dow & 16:02:02.19 & +09:49:07.7 & 57509.47942 & 22.26 & Likely QSO & 1.325$^{+0.205}_{-0.285}$ & 21.05 & 0.23 \\
PS16dov & 16:01:53.91 & +09:52:44.2 & 57508.49664 & 22.25 & QSO & 2.2818$\pm$0.0012 & 21.33 & 0.25 \\
PS16cfu & 16:01:15.66 & +09:25:04.7 & 57508.49664 & 21.14 & Old SN & 0.2809$\pm$0.0655 & 19.25 & 0.28 \\
PS16cgb & 16:02:19.12 & +09:34:50.1 & 57508.49664 & 22.03 & Old SN & 0.7546$\pm$0.1647 & 21.52 & 0.29 \\
PS16cfz & 16:02:11.96 & +09:54:07.9 & 57508.49664 & 21.27 & Old SN & 0.1694$\pm$0.0244 & 17.95 & 0.31 \\
PS16fns & 16:01:24.78 & +10:02:01.6 & 57508.49664 & 21.49 & Likely QSO & 1.955$^{+0.175}_{-0.335}$ & 20.21 & 0.34 \\
PS16cfw & 16:01:39.10 & +09:21:59.5 & 57508.49664 & 21.27 & QSO & 0.8623$\pm$0.0002 & 19.52 & 0.35 \\
PS16fnt & 16:01:28.08 & +10:02:51.0 & 57508.49664 & 21.10 & Likely QSO & 0.975$^{+0.465}_{-0.145}$ & 20.19 & 0.35 \\
PS16fnu & 15:59:41.19 & +09:46:19.7 & 57508.49664 & 22.56 & Stellar & $-$ & 21.41 & 0.37 \\
PS16fnv & 15:59:41.54 & +09:47:50.1 & 57508.49664 & 21.11 & Likely QSO & 1.575$^{+0.205}_{-0.155}$ & 20.45 & 0.38 \\
PS16fnw & 16:01:53.38 & +10:05:21.7 & 57508.49664 & 21.59 & QSO & 2.2851$\pm$0.0004 & 20.04 & 0.42 \\
PS16fnx & 15:59:28.46 & +09:51:51.7 & 57508.49664 & 22.70 & Old SN & 0.5790$\pm$0.1304 & 22.27 & 0.45 \\
PS16fny & 16:02:27.75 & +10:03:09.1 & 57508.49664 & 22.11 & Likely QSO & 0.375$^{+0.135}_{-0.135}$ & 21.31 & 0.47 \\
PS16fnz & 16:01:19.78 & +10:12:18.7 & 57508.49664 & 21.72 & Likely QSO & 0.225$^{+0.245}_{-0.065}$ & 20.69 & 0.51 \\
\hline
\label{tab:ps1}
\end{tabular}
\begin{flushleft}
$^{a}$ Spectroscopic redshifts for QSOs and photometric redshifts for likely old SNe are from SDSS DR12 \citep{alam15}. Photometric redshifts for likely QSOs are from \citet{richards09}.\\
$^{b}$ Angular distance to the final revised coordinates of IceCube-160427A.
\end{flushleft}
\end{table*}

\subsection{Follow-up of PS16cgx}

\begin{figure}
\includegraphics[width=\columnwidth]{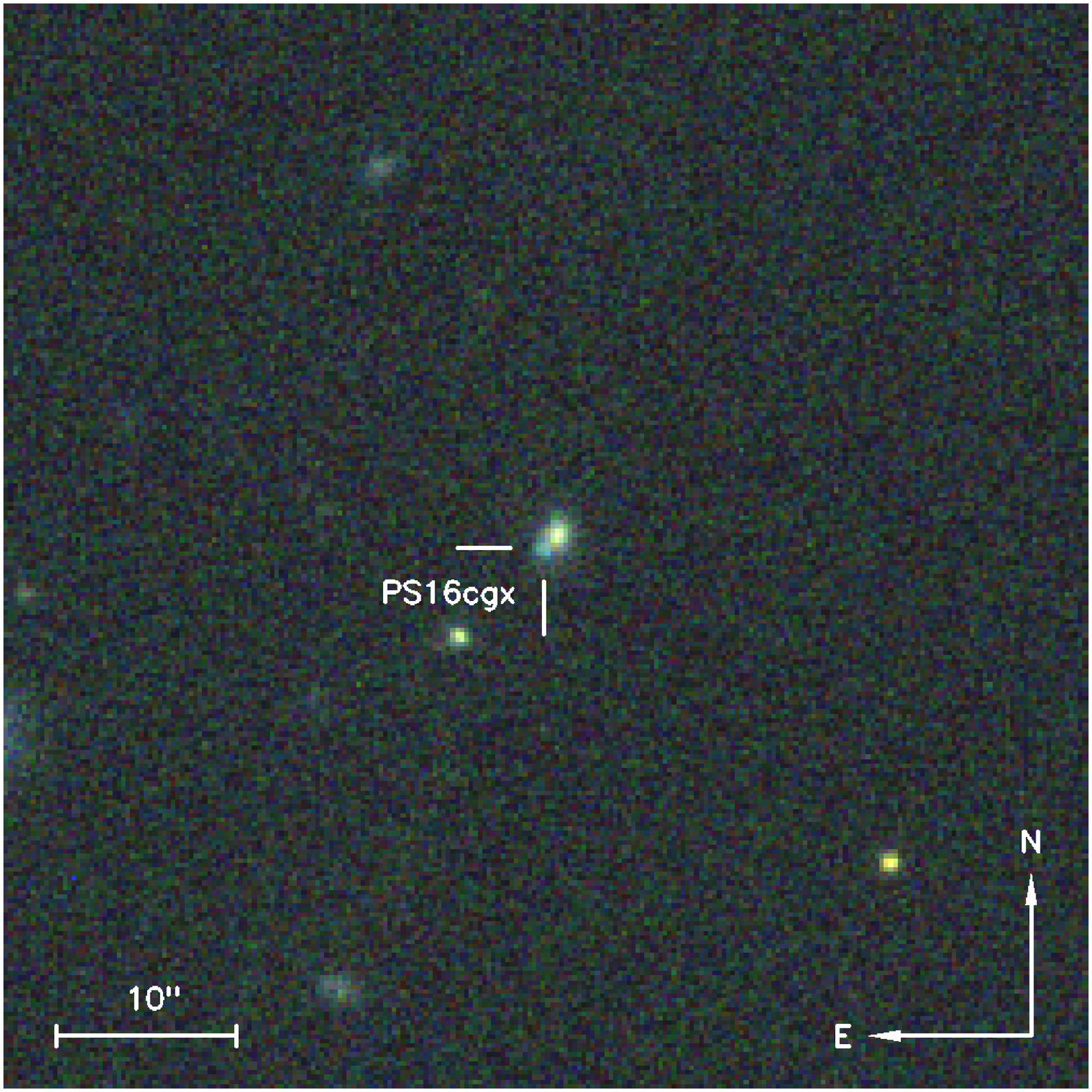}
\caption{Pan-STARRS1 \textit{riz}-band false-colour 1$\arcmin$ $\times$ 1$\arcmin$ subsection of the field of PS16cgx at MJD 57525.5. North is up, east is left.}
\label{fig:field}
\end{figure}

The field of PS16cgx is shown in Fig.~\ref{fig:field}. We continued to observe the field with Pan-STARRS1 until MJD=57550, or 45 days after discovery of the neutrino. These follow-up images were primarily carried out in the \ips-band and the nightly stacked images of the field typically reached $22\lesssim m_{i_{\rm{P1}}} \lesssim 22.5$ (AB mag, $5\sigma$). One epoch of \rps and two epochs of \zps-band imaging were also obtained. Furthermore, two epochs of \textit{i}-band late-time imaging were obtained with the Canada France Hawaii Telescope (CFHT), using the MegaCam instrument \citep{boulade03}. The automatically calibrated level 2 data products were downloaded for the analysis. The CFHT observations were obtained at MJD 57575.3 (2016 July 6.3 UT) suggesting $m_{i} = 23.649 \pm 0.153$ mag, and MJD 57637.2 (2016 September 6.2 UT), with the latter epoch used as a reference template image for the first CFHT epoch. However, it is possible that the latter epoch contains some contamination from the SN. Assuming that PS16cgx is a Type Ia SN, we estimate the magnitude of the SN to be $m_{i} \approx 24.5$ mag during this epoch. Adding this contribution mathematically to the template subtracted magnitude at the MJD 57575.3 epoch results in a magnitude of $m_{i} = 23.2$ mag. The resulting photometry of PS16cgx is reported in Table~\ref{tab:phot}.

Spectra of PS16cgx (see Table~\ref{tab:spect}) were obtained with the following instruments: the 2.2-m University of Hawaii (UH) telescope with SuperNovae Integral Field Spectrograph \citep[SNIFS;][]{lantz04}, the 8.1-m Gemini-North Telescope with Gemini Multi-Object Spectrograph \citep[GMOS;][]{hook04}, and the 8.2-m Very Large Telescope (VLT) with FOcal Reducer and low dispersion Spectrograph 2 \citep[FORS2;][]{appenzeller98}. The SNIFS spectrum was reduced using a dedicated calibration pipeline for the instrument \citep[see also][Section 2 for details of Pan-STARRS follow-up with SNIFS]{smartt16a}. The GMOS spectrum was reduced using the gemini IRAF\footnote{IRAF is distributed by the National Optical Astronomy Observatory, which is operated by the Association of Universities for Research in Astronomy (AURA) under a cooperative agreement with the National Science Foundation.} package. The FORS2 spectra were processed using standard IRAF tasks and reduction steps, and combined for higher signal-to-noise ratio. The redshift of PS16cgx is constrained by the narrow host galaxy emission lines of H$\alpha$ and [S {\sc ii}] 6716.4,6730.8 \AA\ to $z = 0.2895 \pm 0.0001$. This yields a luminosity distance of $D_{l} = 1490$ Mpc ($\mu = 40.87$). The Galactic extinction towards PS16cgx is $A_{i} = 0.073$ mag based on the dust map calibration of \citet{schlafly11}, which is also the calibration system we adopt for all comparisons. Unfortunately, the Na {\sc i} D feature at the rest wavelength of PS16cgx is redshifted in our observer frame to overlap with the heavily contaminating telluric A band around 7615 \AA, and cannot be used to estimate the host galaxy reddening. For the analysis no host galaxy extinction is adopted. The \textit{K}-corrections for the PS16cgx photometry are calculated from the GMOS spectrum with the SNAKE code \citep{inserra18} yielding $K_{rg} = -0.32$, $K_{ir} = -0.28$, and $K_{zi} = -0.29$ mag.

\begin{table}
\caption{Photometry of PS16cgx.}
\begin{tabular}{lllcc}
\hline
\hline
\multicolumn{1}{c}{Epoch} & \multicolumn{1}{c}{$m$} & \multicolumn{1}{c}{$m_{\mathrm{err}}$} & Band & Telescope \\
\multicolumn{1}{c}{(MJD)} & \multicolumn{1}{c}{(mag)} & \multicolumn{1}{c}{(mag)} & &  \\
\hline
57525.48829 & 21.7502 & 0.1878 & \textit{r} & PS1 \\
57508.49664 & 21.8533 & 0.0903 & \textit{i} & PS1 \\
57510.47903 & 21.9087 & 0.3446 & \textit{i} & PS1 \\
57511.46505 & 21.5033 & 0.1409 & \textit{i} & PS1 \\
57512.49131 & 21.6155 & 0.1726 & \textit{i} & PS1 \\
57520.39154 & 21.3547 & 0.0832 & \textit{i} & PS1 \\
57521.47973 & 21.3772 & 0.0731 & \textit{i} & PS1 \\
57523.42637 & 21.3453 & 0.0990 & \textit{i} & PS1 \\
57525.49056 & 21.6267 & 0.1527 & \textit{i} & PS1 \\
57527.47447 & 21.5686 & 0.1447 & \textit{i} & PS1 \\
57528.42873 & 21.6453 & 0.3949 & \textit{i} & PS1 \\
57532.26348 & 22.0882 & 0.4259 & \textit{i} & PS1 \\
57537.38608 & 21.9893 & 0.0751 & \textit{i} & PS1 \\
57540.37896 & 22.0893 & 0.0954 & \textit{i} & PS1 \\
57542.41169 & 22.0684 & 0.1335 & \textit{i} & PS1 \\
57544.41986 & 22.2661 & 0.0745 & \textit{i} & PS1 \\
57546.37147 & 22.1868 & 0.0942 & \textit{i} & PS1 \\
57548.46014 & 22.2848 & 0.1472 & \textit{i} & PS1 \\
57550.46312 & 22.3504 & 0.1179 & \textit{i} & PS1 \\
57575.34358 & 23.649 & 0.153 & \textit{i} & CFHT \\
57525.49283 & 22.2165 & 0.4003 & \textit{z} & PS1 \\
57537.41392 & 22.8827 & 0.3978 & \textit{z} & PS1 \\ 
\hline
\label{tab:phot}
\end{tabular}
\end{table}

\begin{figure*}
\includegraphics[width=2.0\columnwidth]{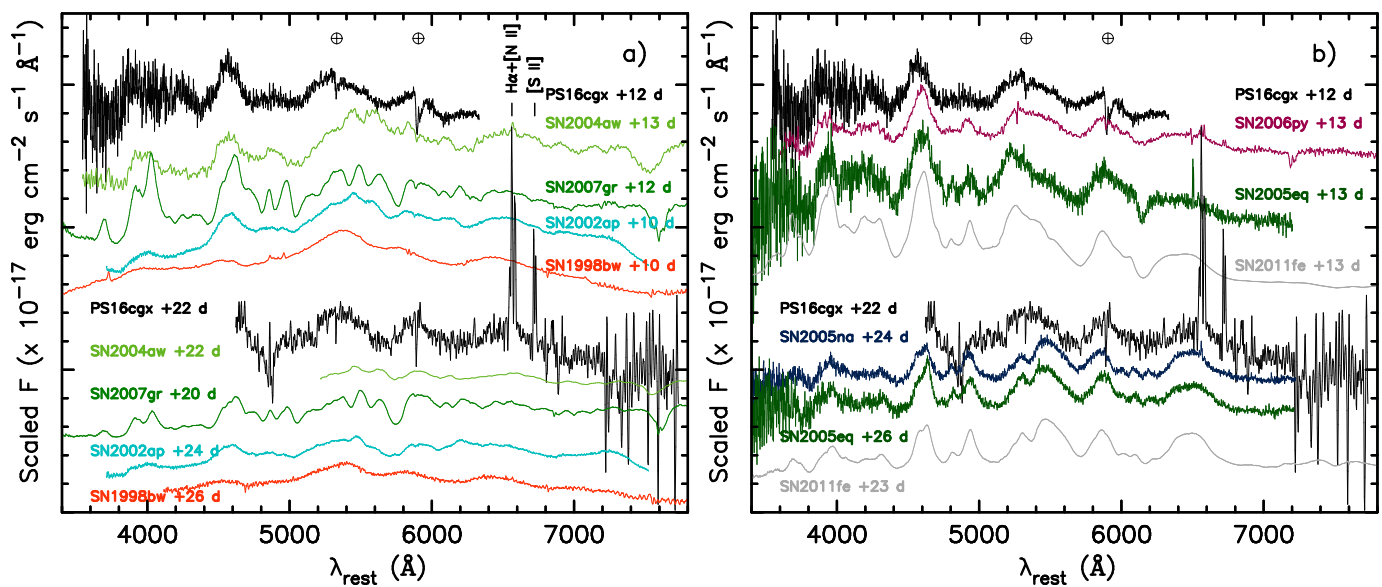}
\caption{a) Spectra of PS16cgx compared to a selection of Type Ic SNe. The comparison events include the GRB associated Type Ic-BL event SN 1998bw \citep{patat01}; non-GRB associated Type Ic-BL event SN 2002ap \citep{modjaz14}; and normal Type Ic events, SN 2007gr \citep{hunter09}, and SN 2004aw \citep{taubenberger06}. The strong host galaxy emission lines of PS16cgx are labelled. b) Spectra of PS16cgx compared to a selection of Type Ia SNe. The comparison events include SN 2005eq, SN 2005na \citep{blondin12}, SN 2006py \citep{folatelli13}, and SN 2011fe \citep{pereira13}. In both panels the spectra have been corrected for line-of-sight reddening and to the rest-frame wavelength. The wavelengths of the main telluric bands in the spectra of PS16cgx are indicated with a $\oplus$ symbol. Comparison data were obtained via WISeREP \citep{yaron12}.}
\label{fig:spect}
\end{figure*}

\begin{table}
\caption{Spectroscopy of PS16cgx.}
\begin{tabular}{ccccc}
\hline
\hline
Epoch & $t_{\mathrm{rest}}$ & Setup & Exp. time & Telescope \\
(MJD) & (d) & & (s) & \\
\hline
57515.49 & $-3.0$ & R-channel & 2$\times$2420 & 2.2-m UH \\ 
57535.42 & $+12.4$ & R400+1.0\arcsec & 4$\times$900 & Gemini-N \\ 
57547.19 & $+21.6$ & 300I+1.0\arcsec & 2700 & VLT-UT1\\ 
57548.18 & $+22.3$ & 300I+1.0\arcsec & 2$\times$2400 & VLT-UT1 \\ 
\hline
\label{tab:spect}
\end{tabular}
\end{table}

\subsection{Nature of PS16cgx}

Due to the faint observed brightness of PS16cgx around 21 to 22 mag in \textit{i}-band, it is challenging to obtain high signal-to-noise spectra of such an event even with an 8-m class telescope. Due to this, the nature of PS16cgx is ambiguous between a Type Ia and a Type Ic SN based solely on spectroscopic classification (Fig.~\ref{fig:spect}). A Type Ic origin of PS16cgx would be extremely interesting, since both GRB associated Type Ic SNe and also some normal Type Ic SNe \citep[e.g.][]{paragi10} are associated with relativistic jets immediately after the core collapse. These jets are expected to produce a short ($\sim$10 s) burst of neutrinos; a detection of such neutrinos could produce valuable insights into the physics of these jets \citep[e.g.][]{ando05} and identify the class of choked-jet SNe as neutrino emitters. 

The lack of very broad features in the spectra of PS16cgx suggest that it is not a broad-line Type Ic (Type Ic-BL) SN such as SN1998bw \citep{patat01}. While the spectra of PS16cgx do not exactly resemble those of the normal Type Ic SN 2007gr \citep{hunter09} either, there is some similarity for example with the Type Ic SN 2004aw \citep{taubenberger06}. The main difference in comparison of PS16cgx to normal Type Ia SNe is the shallow absorption feature that could be associated with the \SiII\ line. Nonetheless, the early spectrum of PS16cgx shares significant similarity to that of Type Ia SNe, such as SN 2006py \citep{folatelli13}.

\begin{figure*}
\includegraphics[width=2.0\columnwidth]{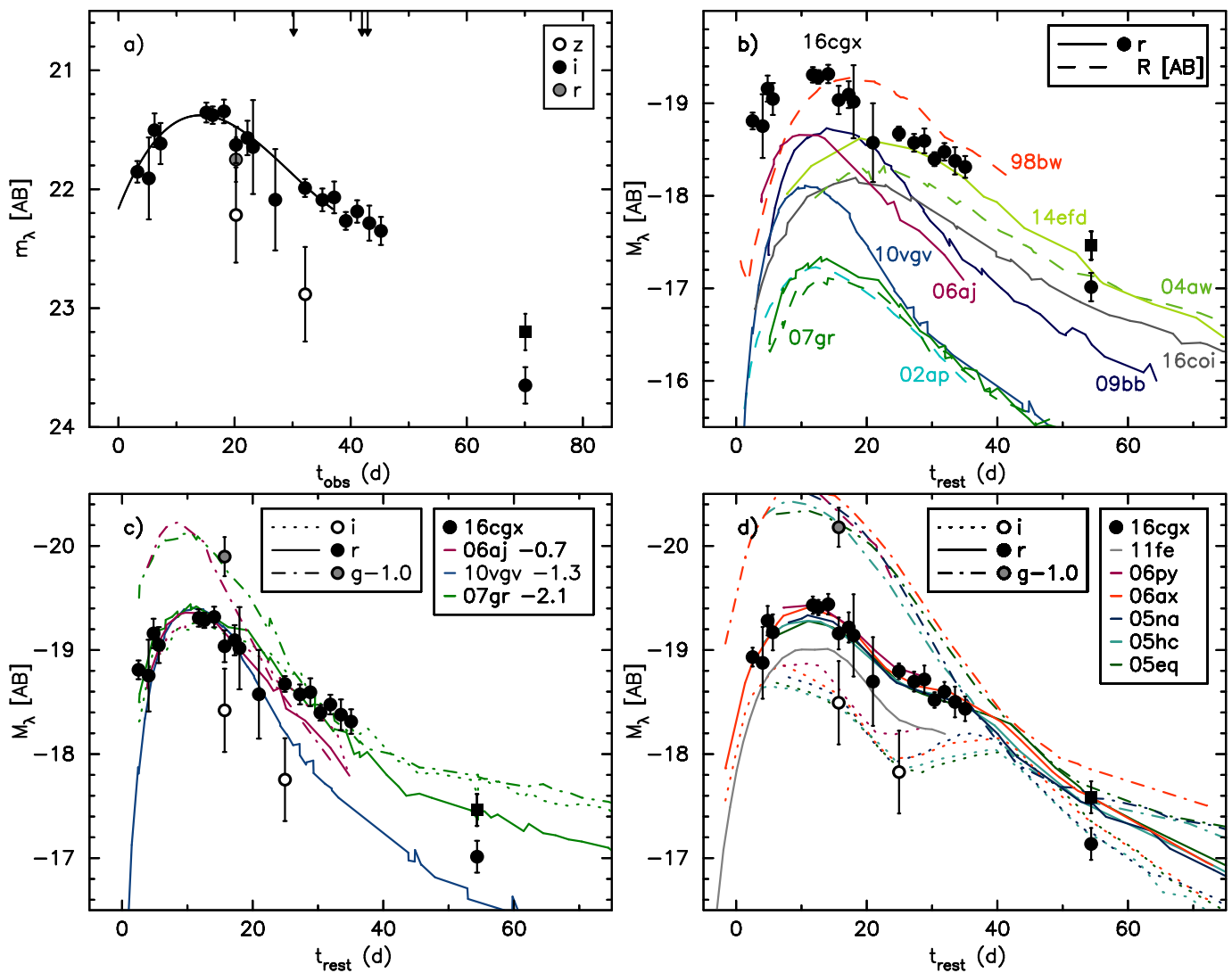}
\caption{a) Observed light curves of PS16cgx. Zero epoch is set to the IceCube-160427A detection. A low-order polynomial light-curve fit is shown with a solid curve. Spectroscopic GMOS (30 days) and FORS2 (42 and 43 days) observations are indicated with vertical arrows. b) A PS16cgx \textit{r}-band absolute light-curve comparison to a selection of nearby Type Ic SNe including the GRB associated Type Ic-BL events (SNe 2006aj and 1998bw); non-GRB associated Type Ic-BL events (PTF10vgv, and SNe 2016coi, 2009bb, and 2002ap); and normal Type Ic events (LSQ14efd, and SNe 2007gr and 2004aw). c) A selection of rapidly rising Type Ic SNe arbitrarily shifted as indicated in the figure legend. The comparison indicates an inconsistent spectrophotometric evolution of PS16cgx with Type Ic SNe which cannot be explained with extinction effects. d) A PS16cgx \textit{gri} absolute light-curve comparison to a selection of similar Type Ia SNe. In addition, the slightly fainter \textit{r}-band light curve of the canonical Type Ia SN 2011fe \citep{guillochon17} is also included in the comparison. Furthermore, as a difference to the previous panels, the light curves of PS16cgx have been arbitrarily corrected for a modest host galaxy extinction of $A_{V} = 0.2$ mag with $R_{V} = 1.27$ derived by \citet{burns14} as an average value for their Type Ia sample. Assuming that the CFHT template reference image contains contamination from a Type Ia SN at $m_{i} = 24.5$ mag, the final photometry point is also shown corrected for this effect with a square symbol in all panels.
\label{fig:phot}}
\end{figure*}

Based on a third order polynomial fit to the Pan-STARRS1 \ips-band observations near maximum light, we estimate that the PS16cgx light curve peaked at 11 rest-frame days after the neutrino detection (corresponding to MJD 57519.4) and at $M_{r} \approx -19.3$ mag (Fig.~\ref{fig:phot}a). Such an approach is often used to estimate the peak magnitudes of a SN \citep[e.g.][]{suntzeff99}. Compared to the absolute peak magnitude ranges (corrected for host galaxy line-of-sight extinction) yielded by \citet{drout11}, PS16cgx is brighter than normal Type Ic SNe with a peak magnitude of $M_{R} = -18.1 \pm 0.6$ mag, but consistent with Type Ic-BL SNe peaking at $M_{R} = -18.8 \pm 1.1$ mag (\textit{R}-band magnitudes scaled to AB system). For comparison, \citet{taddia18} obtain an average peak magnitude of $M_{r} = -17.66 \pm 0.21$ mag for Type Ic SNe. This is also reflected in our comparison of the absolute light curve of PS16cgx to a variety of nearby Type Ic SNe with well estimated explosion dates based on observational constraints or modelling (Fig.~\ref{fig:phot}b). The comparison events include the GRB associated Type Ic-BL events SN 2006aj \citep{bianco14} and SN 1998bw \citep{galama98}; non-GRB associated Type Ic-BL events SN 2016coi \citep{prentice18}, PTF10vgv \citep{corsi12}, SN 2009bb \citep{pignata11}, and SN 2002ap \citep{foley03}; and normal Type Ic events LSQ14efd \citep{barbarino17}, SN 2007gr \citep{hunter09}, and SN 2004aw \citep{taubenberger06}. The Johnson-Cousins light curves of SN 2007gr are also converted to the SDSS system using the transformations of \citet{jester05} in the comparison. Virgo and Great Attractor corrected host luminosity distances (via the NASA/IPAC Extragalactic Database) are used in comparisons if relevant. 

Furthermore, most of the Type Ic SNe in our comparisons have slower rise times than PS16cgx would have to have to be associated with the IceCube-160427A event (assuming that neutrino emission is temporally coincident with core collapse or shock breakout). While some of the normal Type Ic SN light curves could be arbitrarily shifted to match the overall shape of the rest-frame \textit{r}-band evolution of PS16cgx, the \textit{r}$-$\textit{i} colour evolution of events like SN 2006aj \citep{bianco14} and SN 2007gr \citep{hunter09} do not appear to be consistent with that of PS16cgx (Fig.~\ref{fig:phot}c). 

\begin{figure*}
\includegraphics[width=\columnwidth]{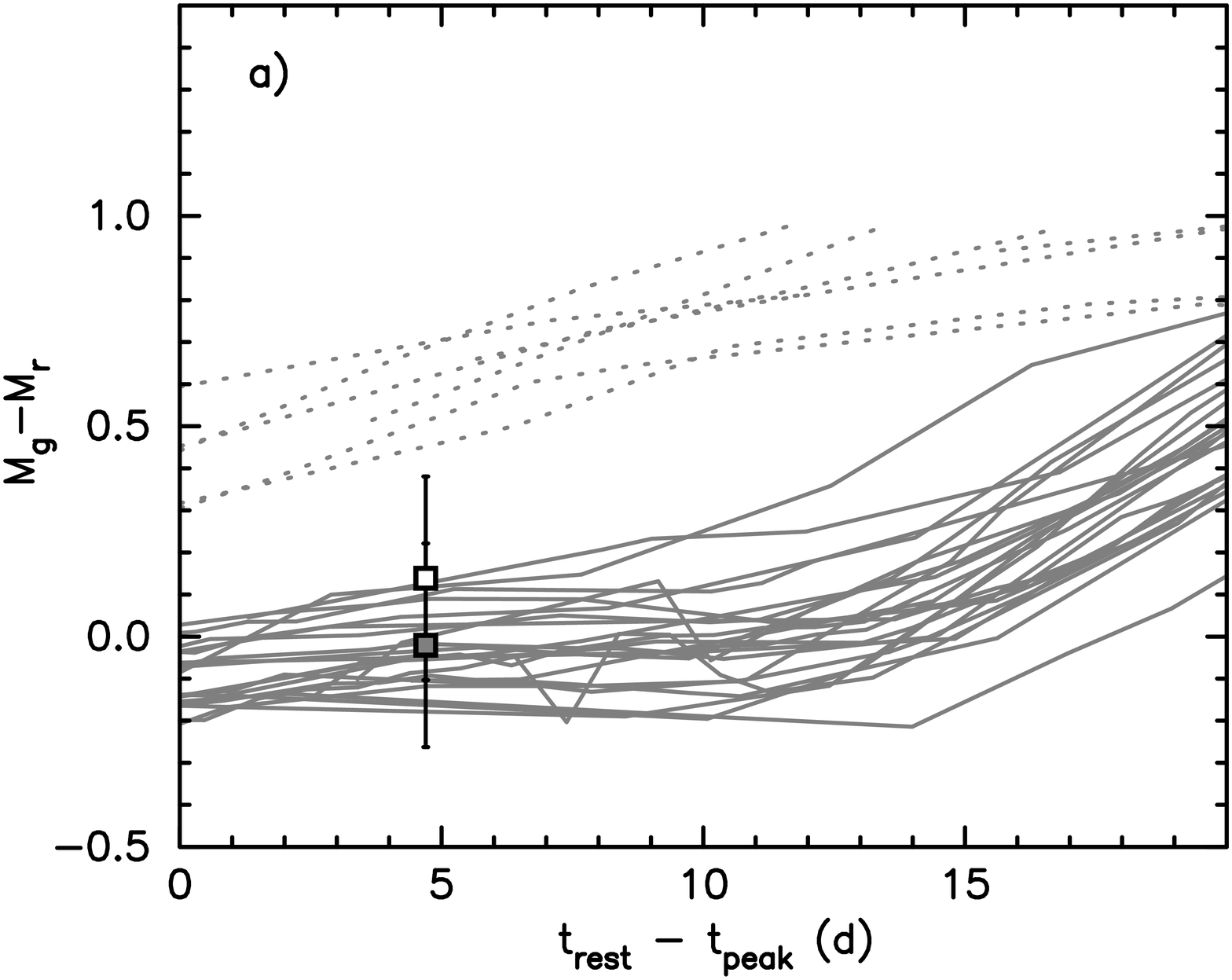}
\includegraphics[width=\columnwidth]{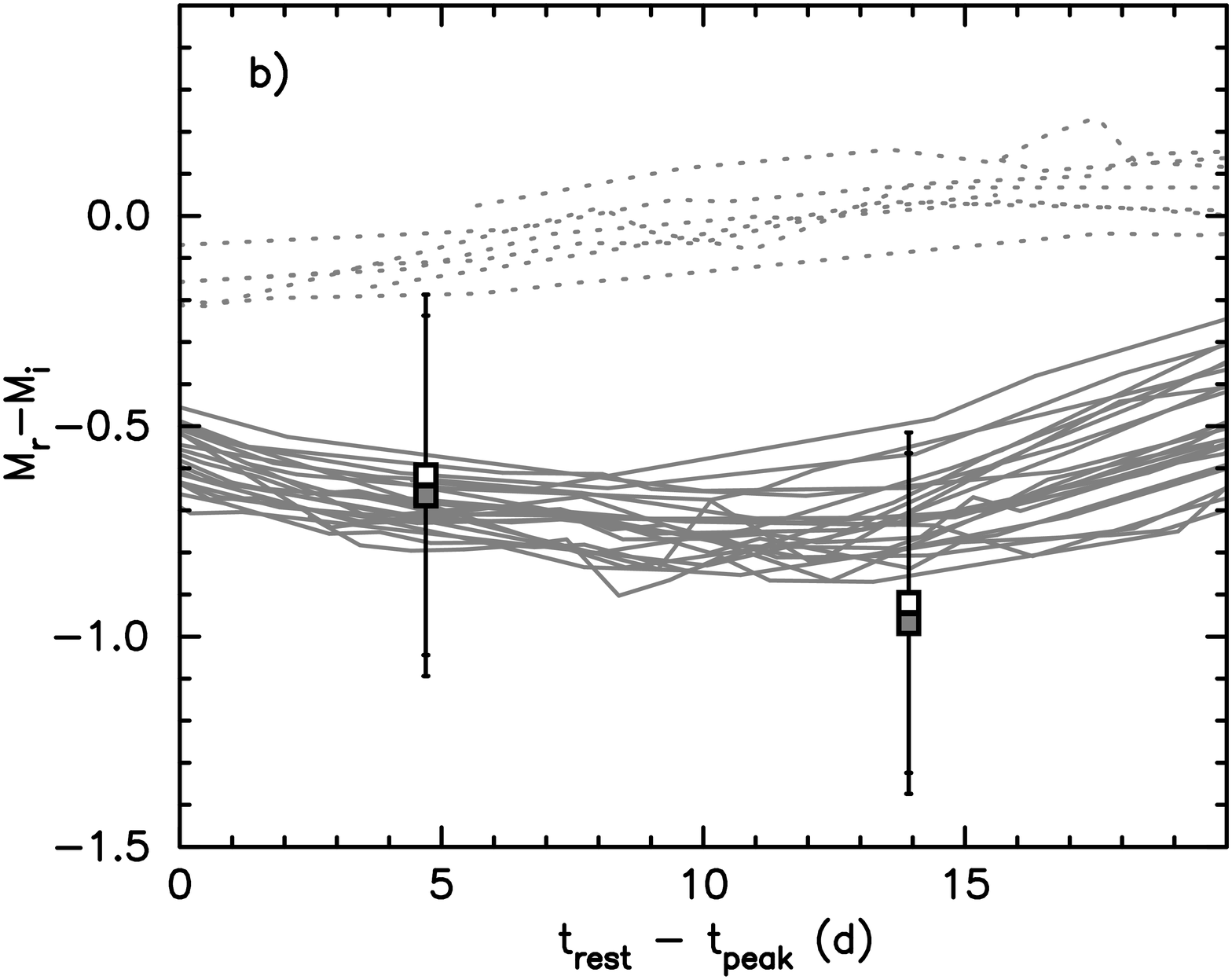}
\caption{a) Rest-frame $M_{g} - M_{r}$ and b) $M_{r} - M_{i}$ colours of PS16cgx compared to the evolution of Type Ia SNe \citep[solid curves;][]{burns14,krisciunas17} including both normal and 91T-like events, and normal Type Ic SNe \citep[dotted curves;][]{stritzinger18a,stritzinger18b,taddia18} from CSP. Colour points assuming no host galaxy extinction are shown with open symbols, and assuming an arbitrary host extinction of $A_{V} = 0.2$ mag in grey symbols.
\label{fig:colour}}
\end{figure*}

Contrary to the Type Ic SN comparisons, the absolute light curve evolution of PS16cgx is similar to that of many normal Type Ia SNe (Fig.~\ref{fig:phot}d), for example SNe 2005eq, 2005hc, 2005na, 2006ax, and 2006py from the Carnegie Supernova Project \citep[CSP;][]{contreras10, burns14, krisciunas17}. In particular, this is the case, if a small host galaxy extinction is assumed. Another approach to illustrate this is to compare the rest-frame colour curves of Type Ia and Ic SNe to the observations of PS16cgx (Fig.~\ref{fig:colour}); this comparison suggests that the colours of PS16cgx are consistent Type Ia SNe, but bluer than normal Type Ic SNe. This discrepancy with Type Ic events cannot be explained with an underestimated host galaxy extinction of PS16cgx. Furthermore, while multi-band observations of PS16cgx are consistent with Type Ia SNe in the literature within errors, the match would be improved assuming a moderate host galaxy extinction of around $A_{V} \approx 0.2$ mag.

To further quantify the compatibility of PS16cgx with the Type Ia SN hypothesis, we fit the $riz$-band light curve with the 4-parameter SALT2 template \citep{guy05,guy07}, assuming a Galactic extinction of $E(B-V) = 0.054$ mag derived from \citet{finkbeiner99} and a host redshift of $z = 0.2895$. The best-fit parameters were $t_0 = 57517.55 \pm 0.59$~(MJD, i.e. 2016 May 9.15 UT) for peak, $x_0 = (4.39 \pm 0.71) \times 10^{-5}$, $x_1 = -0.48\pm0.61$, and $c = -0.04\pm0.13$. The allowed range for the $x_0$ parameter corresponds to a distance modulus of $\mu = 40.89 \pm 0.18$, perfectly compatible with the $\Lambda$CDM prediction of $40.89$ at $z=0.2895$. Similarly, the $x_1$ (stretch) and $c$ (colour) parameters are compatible with the mean values (0) for Type Ia SNe. Furthermore, we studied the impact of colour information by testing the agreement between the data and template in each band. Since 19 of the 22 observations were taken in the $i$ band, this band anchors the fit. However, the photometric points in both the $r$ and $z$ band fall slightly below the model expectation, which is an effect that cannot be explained with a simple reddening law. Quantitatively, the $\chi^2$/d.o.f. for the fit is 19.7/18 ($p = 0.35$), and the $riz$ light curve can be considered compatible with the template to within photometric errors. To summarize, we conclude that there are no specific arguments to support a Type Ic classification of PS16cgx based on the available data. However the observations are in reasonable agreement with the expected Type Ia SN light curve and colour evolution. While some theoretical models~\citep{wright16,wright17} predict that MeV neutrinos from a nearby Type Ia SN would be detectable by IceCube, no neutrino emission is expected at TeV energies and above. Therefore, we find no plausible physical scenario where the neutrino event could be related to PS16cgx if it were indeed a Type Ia SN and conclude that in that case the two detections must be unrelated.

The host galaxy SDSS J160118.56+095153.7 of PS16cgx is an extended source in our images, however, no clear structure is resolved. Based on the Pan-STARRS1 DR1, the host has \textit{K}-corrected absolute magnitudes of $M_{g} \approx -20.7$ and $M_{r} \approx -21.2$ mag. The luminosity-metallicity relation of \citet{tremonti04} suggests $12 + \log(\mathrm{O}/\mathrm{H}) \approx 9$. The metallicity using \citet{pettini04} N2 method is $12 + \log(\mathrm{O}/\mathrm{H}) \approx 8.7$ yielded from the VLT spectrum, consistent with the mass-luminosity relation. Therefore, the host has relatively normal, solar-like, metallicity. 

\subsection{Probabilities}
\label{sec:prob}
The probability of a chance detection of PS16cgx with the IceCube-160427A event can be estimated following a similar approach to that of \citet{aartsen15} and \citet{smartt16b}. Both cases of a Type Ia or Type Ic SN are considered. Assuming that a young and very distant SN can be discovered and that it can be identified to be in rise, the detection would likely have to take place roughly 0.5 mag below the peak. The brightest events we consider are the Type Ic-BL SNe which could peak roughly at $M = -22.1$ mag assuming a 3$\sigma$ deviation from an average magnitude \citep{drout11}. Therefore, with the Pan-STARRS1 survey limit of $\sim$22.5 mag, such events could be discovered at rise up to a luminosity distance of $D_{l} \approx 6610$ Mpc (or $z\approx 1.00$). However, the majority of the events within this volume are too faint to be discovered. The review on the cosmic star formation rate by \citet{madau14} yields a CCSN rate of $R_{\mathrm{CCSN}} = 2.0 \times 10^{-4}$ Mpc$^{-3}$ yr$^{-1}$ scaled to the redshift of PS16cgx, which very roughly corresponds also to a comoving mid-volume distance from within an average magnitude Type Ic-BL or a normal Type Ia SN could be discovered with this survey. We adopt this CCSN rate as a rough average for the calculations. An intrinsic number of exploding CCSNe can be roughly estimated from

\begin{equation}
N_{\mathrm{SN}} = \frac{4}{3} \pi D_{m}^{3} R_{\mathrm{SN}} \frac{A}{41253~\mathrm{deg}^{2}} \frac{t}{365~\mathrm{d}} \mathrm{yr,}
\end{equation}

where $D_{m}$ is the comoving radial distance $D_{m} = D_{l}/(1+z)$. With a sky area given by the neutrino error region of $A = 0.78 $ deg$^{2}$ and a survey time window of $t=5$ d, this suggests $N_{\mathrm{CCSN}} = 7.8$. For the fractions of normal Type Ic and Type Ic-BL SNe from all CCSNe in a volume-limited sample, \citet{li11} derive $f_{\mathrm{Ic}} = 14$\% and $f_{\mathrm{Ic-BL}} = 2$\%, respectively \citep[see also][]{smith11}, suggesting $N_{\mathrm{Ic}} = 1.1$ and $N_{\mathrm{Ic-BL}} = 0.16$. To estimate which fraction of the events are bright enough to be discovered with the survey, two effects have to be taken into account. Firstly, \citet{mattila12} estimates that only 77\% of CCSNe at the redshift of PS16cgx are observable by optical surveys. This is due to a missing fraction of CCSNe with very high localised host galaxy extinctions way above the `normal' extinction distribution in high star formation rate environments. Such correction reduces the discrepancy between the directly observed CCSN rates and the predicted star formation rates \citep{horiuchi11}, and has been applied by CCSN rate studies \citep[e.g.][]{dahlen12,melinder12}. Secondly, a transient that is possibly associated with a neutrino detection is a young SN discovered at $>$0.5 mag below the following peak magnitudes: $M_{\mathrm{Ic}} = -18.1$ mag and 1$\sigma$ deviation $\sigma_{\mathrm{Ic}} = 0.6$ mag, or $M_{\mathrm{Ic-BL}} = -18.8$ mag and $\sigma_{\mathrm{Ic-BL}} = 1.1$ mag \citep{drout11}. Furthermore, we assume a normal rest-frame \textit{r}-band host extinction distribution for SNe based by \citet{riello05} and scaled as in \citet{mattila12}. Using the aforementioned limiting magnitude of $\sim$22.5 mag, this yields a detection of only 1.4\% of normal Type Ic SNe within the comoving volume. Combining the detection and missing fraction corrections results in an observable $N_{\mathrm{Ic}} = 0.012$, corresponding to a Poisson probability $p = 1 - e^{-N}$ of $p_{\mathrm{Ic}} = 1.2$\%. Similarly, a normal detection efficiency of 4.7\% can be yielded for Type Ic-BL SNe, which results in an observable $N_{\mathrm{Ic-BL}} = 0.006$ and $p_{\mathrm{Ic-BL}} = 0.6$\% for a chance detection.

The Type Ia SN rate at the redshift of PS16cgx is $R_{\mathrm{Ia}} = 0.4 \times 10^{-4}$ Mpc$^{-3}$ yr$^{-1}$ \citep{dilday10}. For the aforementioned large comoving volume this suggests $N_{\mathrm{Ia}} = 1.6$. The host galaxy extinction distribution for Type Ia SNe is adopted from \citep{dilday10}. No additional missing fraction is assumed for Type Ia SNe that are rising from old stellar population. Luminosity functions at peak for Type Ia SNe are adopted from \citet{melinder11} assuming \textit{B}$-$\textit{R} $=0$ mag and shifted to the AB system (see Table~\ref{tab:prob}). These result in detection efficiencies at $>$0.5 mag below peak of 4.7, 6.1, 0.9, and 3.4\% for normal, 91T-like, 91bg-like, and faint Type Ia SNe, respectively. This yields a total detection efficiency of 4.4\% within the volume and thus a number of detectable Type Ia SNe of $N_{\mathrm{Ia}} = 0.070$ corresponding to a chance detection of $p_{\mathrm{Ia}} = 7$\%. Therefore, a chance detection of a young and unrelated Type Ia SN in the Pan-STARRS1 follow up of an IceCube neutrino alert sky region is not that unlikely. The adopted and yielded values of all the above probability calculations including different Type Ia subtypes are shown for clarity in Table~\ref{tab:prob}.

\begin{table*}
\centering
\caption{Adopted and yielded values of the probability calculations of a chance discovery for the IceCube-160427A error region.}
\begin{tabular}{ccccccccc}
\hline
\hline
Type & $f_{\mathrm{relative}}$ & $N_{\mathrm{intrinsic}}$ & $f_{\mathrm{missing}}$ & $M_{\mathrm{peak, AB}}$ & $\sigma$ & $f_{\mathrm{detection}}$ & $N_{\mathrm{observed}}$ & $p$ \\
 & (\%) & & (\%) & (mag) & (mag) & (\%) & & (\%) \\
\hline
CCSN & 100 & 7.8 & 77 & - & - & - & - & - \\
Ic & 14 & 1.1 & 77 & $-18.1$ & 0.6 & 1.4 & 0.012 & 1.2 \\
Ic-BL & 2 & 0.16 & 77 & $-18.8$ & 1.1 & 4.7 & 0.006 & 0.6 \\
\hline
Ia & 100 & 1.6 & - & - & - & 4.4 & 0.070 & 6.8 \\
Ia-normal & 62 & 0.99 & - & $-19.1$ & 0.5 & 4.7 & 0.047 & 4.6 \\
Ia-91T & 12 & 0.19 & - & $-19.4$ & 0.3 & 6.1 & 0.012 & 1.2 \\
Ia-91bg & 6 & 0.10 & - & $-17.6$ & 0.5 & 0.9 & 0.001 & 0.1 \\
Ia-faint & 20 & 0.32 & - & $-18.8$ & 0.5 & 3.4 & 0.011 & 1.1 \\
\hline
\label{tab:prob} 
\end{tabular}
\end{table*}

\section{Summary of other Pan-STARRS1 observations of IceCube alerts}
\label{sect:summ}

Over the two-year period 2016 April 1 -- 2018 March 31, IceCube released 13 alerts of either EHE or HESE events. Of these thirteen, ten were visible in the night time sky (the other three were in solar conjunction) and with Pan-STARRS1 we observed the field of 5 of them (see Table\,\ref{tab:summ} for a summary). 
Since usually the initially released neutrino direction changes slightly after application of more time consuming reconstructions, we will use the latest available reconstruction results as our final set of coordinates. We typically observed with a similar observing strategy as discussed here in Section\,\ref{sect:Supernova} with dithered exposures stacked together to make a night stack (as for the Pan-STARRS1 Medium Deep Fields).  The size of the Pan-STARRS1 camera field of view and typical sky localisation regions from the EHE and HESE tracks from IceCube are well matched and we typically covered the neutrino error region to a depth of $m_{i_{\rm{P1}}} \approx 22$ mag. Observations of the field started between a few hours to a few days after the IceCube alert and continued for 3 to 30 days afterwards. We were motivated to look both for rapidly evolving transients (timescales of order 24$-$48 hrs) and SN-like transients (timescales of order a few to 10 days). As listed in Table\,\ref{tab:summ}, we typically found 10$-$20 transients per field, with magnitude ranges and light curves broadly similar to those detailed for IceCube-160427A in Section~\ref{sect:Supernova}. However, we found no clear candidate for a SN-like transient with a rising light curve (or indeed any other characteristic) that would point towards a temporal coincidence. PS16cgx was the best candidate for a SN with an explosion epoch coincident with the neutrino time, and we consider that to be disfavoured due to its likely thermonuclear origin. The numbers of transients that we found are entirely consistent with the typical SN field population in a Pan-STARRS1 footprint at this depth (the vast majority are old SNe) and variable QSOs and AGNs (see Appendix\,\ref{sec:app}). 

We cannot make a definitive statement that all transients discovered can be discounted as counterparts. We can only say that none of them show any unusual characteristics in their light curves or absolute magnitudes (when a host galaxy redshift exists) or in spectral characteristics (when a spectrum has been taken, see Appendix\,\ref{sec:app}) that suggest they are candidate sources for the counterparts of any of the HESE or EHE neutrinos. In other words, we are confusion limited within the $\sim$1$-$2\,deg$^2$ localisation regions of IceCube.  Given the number of sources detected, we have only the option of searching for an unusual observation characteristic or implied temporal coincidence. Either of these significantly  impact the probability calculation of coincidence as shown in Section\,\ref{sec:prob}. 

We can make a qualitative statement that we find no evidence of transients above $m_{i_{\rm{P1}}} \approx 22$\,mag in the error regions. However the purity of the neutrino alerts is between 25$-$50\% and we have not been able to cover all five events from 1 hour to 10 days uniformly. Therefore we cannot rule out fast transients of timescales $\sim$minutes to 24 hrs or fainter transients with longer timescales. 

\section{Discussion}
\label{sect:Discussion}

\begin{figure}
\centering
\includegraphics[width=\columnwidth]{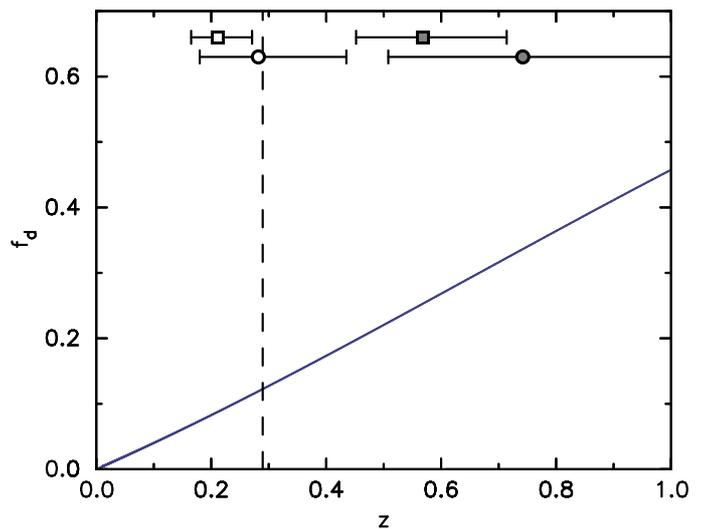}
\caption{Cumulative fraction, $f_d$, of astrophysical neutrino sources with a single high-energy neutrino detection in IceCube, accumulated as a function of source redshift, assuming they follow the SFR \citep{madau14} given standard cosmology. The redshift of PS16cgx is marked with a vertical line. The maximum distance ranges where a normal Type Ic (squares) and Type Ic-BL (circles) SNe can be detected in rise are indicated assuming negligible line-of-sight extinction, ignoring \textit{K}-corrections, and adopting 1$\sigma$ distribution of the SN peak magnitudes, for image depths of 22.5 and 25 mag (white and grey symbols, respectively).}
\label{fig:singlet_distances}
\end{figure}

The distance distribution of SNe follow the star-formation rate. If we assume those are the sources of the diffuse neutrino signal, the median redshift of sources of high-energy neutrino singlets is at $z \sim 1$. For example, with Pan-STARRS1 image depth of $\sim$22.5 mag typical normal Type Ic and Type Ic-BL SNe can be detected in rise up to $z = $~0.2$-$0.4. However, superluminous Type Ic events could be detected up to $z \approx 1$. Out of the source, which produce a single detectable high-energy neutrino alert in IceCube, only 13\% are closer than redshift of $z = 0.29$ (see Fig.~\ref{fig:singlet_distances}). Finding a distant neutrino source is therefore not surprising. Due to the Eddington bias we cannot estimate the neutrino flux from the detection of a single high-energy neutrino event by using Poisson statistics. As a consequence of the Eddington bias, the detection of a single high-energy neutrino implies an expectation value much smaller than one~\citep{strotjohann19}. Unlike previous searches for neutrino multiplets \citep{abbasi12}, the observation of single neutrinos do not constrain the distance or luminosity of the source very much. Therefore a deep optical follow-up is necessary to catch the counterparts of distant sources. In the following we calculate the probability to find an optical counterpart in the follow-up of single high-energy neutrinos as a function of the survey depth.

We focus on HESE and EHE track-like alerts for which we can assume that their location in the sky will be confined to within 1 deg$^{2}$ uncertainty. The rate of HESE/EHE neutrinos produced in a redshift shell is nearly independent of the redshift up to $z \sim 0.6$ if we assume that the redshift distribution of the sources (SN of some sort) follow the star-formation rate (SFR). Fig.~\ref{fig:singlet_distances} shows the accumulated fraction of sources as a function of redshift. We expect about four HESE and another four EHE events per year, of which on average about one and two, respectively, should be of true astrophysical origin. With a rate of $R_{\nu} = 3$ yr$^{-1}$ astrophysical track-like events we can expect to detect a true association with a rate $R_{\nu} \cdot f_{d} \sim 0.4$ yr$^{-1}$ up to a redshift 0.3 assuming that all IceCube alerts can be followed up.

Accordingly, a deeper search for optical counterparts is well motivated\footnote{We note that his argument holds for sources with positive evolution. In the rare case of negative evolution, for example for tidal disruption events, a deeper search might not be beneficial.}. Extending the redshift range, for instance from 0.3 to 0.6 will nearly double the number of source identifications.

\begin{table*}
\setlength{\tabcolsep}{4pt}
\centering
\caption{IceCube HESE and EHE events and Pan-STARRS1 observations from 2016$-$2017. The events labelled `Not visible' have IceCube positions in the daytime sky. Events marked `Not observed' were either due to weather, Moon encroachment or other reasons not to observe the field. The information on the final reconstructed sky locations of the IceCube neutrinos are taken from the IceCube catalogue of alert events up through IceCube-170922A$^{a}$, or from the \citet{icecube17a,icecube17b,icecube17c}. The number of transients are those within the 90\% error localisation.}
\begin{tabular}{ccccccccccc}
\hline
\hline
Neutrino event & Epoch  & Energy alert & $\alpha_{\mathrm{J2000}}$ & $\delta_{\mathrm{J2000}}$ & 90\% error & \multicolumn{2}{c}{Pan-STARRS1 observations} & $m_{\mathrm{depth},i_{\rm P1}}$ & Number of \\
 & (MJD) & type & ($^{\circ}$) & ($^{\circ}$) & ($^{\circ}$) & Start & End & (mag) & transients \\\hline 
IceCube-160427A & 57505.24481 & HESE & 240.3 & 9.7 & 0.45$-$0.54 & 57508.50 & 57550.46 & 22.5 & 19 \\
IceCube-160731A & 57600.07990 & HESE+EHE & 214.5 & -0.3 & 0.75 &  57600.32 & 57626.28 & 22 & 11 \\ IceCube-160806A & 57606.51496 & EHE & 122.8 & -0.8 & 0.5 & Not visible & - & - & - \\
IceCube-160814A & 57614.90688 & HESE & 200.3 & -32.4 & 1.2$-$3.0 & Not visible & - & - & - \\
IceCube-161103A & 57695.38022 & HESE & 40.8 & 12.6 & 0.65$-$1.1 & 57695.47 & 57702.36  & 22 &  7 \\
IceCube-161210A & 57732.83797 & EHE & 46.6 & 15.0 & 0.4$-$1.1 & Not observed & - & - & - \\
IceCube-170312A & 57824.57615 & HESE & 305.2 & -26.6 & 0.5 & Not observed & - & - & - \\
IceCube-170321A & 57833.31413 & EHE & 98.3 & -15.0 & 1.2 & 57834.27 & 57859.27 & 22 & 2 \\
IceCube-170506A & 57879.52565 & HESE & 221.8 & -26.0 & 2.0$-$3.0 & Not observed & - & - & - \\
IceCube-170922A & 58018.87118 & EHE & 77.4 & 5.7 & 0.3$-$0.95 & Not observed & - & - & - \\
IceCube-171015A & 58041.06562 & HESE & 162.9 & -15.4 & 1.6$-$2.6 & Not visible & - & - & - \\
IceCube-171028A & 58054.35295 & HESE & 67.5 & -69.8 & $>$7$-$28 & Not observed & - & - & - \\
IceCube-171106A & 58063.77754 & EHE & 340.0 & 7.4 & 0.25$-$0.7 & 58065.34 & 58075.34 & 22.5 & 2 \\ 
\hline
\label{tab:summ} 
\end{tabular}
\begin{flushleft}
$^{a}$ \url{https://icecube.wisc.edu/science/data/TXS0506_alerts}
\end{flushleft}
\end{table*}

\section{Conclusion}
\label{sect:Conclusion}

We presented the properties of the first IceCube high-energy starting event distributed to the public, IceCube-160427A. The neutrino direction was observed by various instruments in gamma-ray and optical wavelengths. The only compelling optical counterpart found was PS16cgx. Our detailed investigations show some ambiguity in the classification of this SN. While the optical spectra could be explained by both Type Ia and Ic scenarios, the light curve and colour evolution points to PS16cgx being a Type Ia SNe and therefore likely unrelated to IceCube-160427A. The higher chance probability for Type Ia SNe supports this scenario. Therefore, we report a Pan-STARRS1 follow-up non-detection of an optical counterpart for IceCube-160427A within the 90\% confidence region down to a 5$\sigma$ limit of $m_{i_{\rm{P1}}} \approx 22.5$ mag.

Our findings show that spectroscopic follow-up and multi-colour photometric coverage of the light curves of candidate neutrino sources are crucial. Especially, spectral coverage during the SNe peak would have made a classification easier, since at late times Type Ic and Type Ia spectra look somewhat similar.

We also find no other plausible candidate for an optical transient coincident with four IceCube EHE/HESE triggers down to the 5$\sigma$ limit of $m_{i_{\rm{P1}}} \approx 22$ mag. A large fraction of potential CCSN counterparts of high-energy neutrino detections are not accessible by shallow surveys. Assuming that high-energy neutrinos are produced by transients that trace the star-formation rate and have an optical counterpart, we suggest to extend follow-up observations to surveys reaching redshifts of up to $0.6$ to ensure that the volume encloses a high probability of detection of at least one counterpart per year.

\begin{acknowledgements}

\textbf{Pan-STARRS:} Pan-STARRS is supported by the University of Hawaii and the NASA’s Planetary Defense Office under Grant No NNX14AM74G and the Pan-STARRS Multi-messenger science collaboration is 
supported by Queen’s University Belfast.
The Pan-STARRS1 Surveys (PS1) have been made possible through contributions of the Institute for Astronomy, the University of Hawaii, the Pan-STARRS Project Office, the Max-Planck Society and its participating institutes, the Max Planck Institute for Astronomy, Heidelberg and the Max Planck Institute for Extraterrestrial Physics, Garching, The Johns Hopkins University, Durham University, the University of Edinburgh, Queen's University Belfast, the Harvard-Smithsonian Center for Astrophysics, the Las Cumbres Observatory Global Telescope Network Incorporated, the National Central University of Taiwan, the Space Telescope Science Institute, the National Aeronautics and Space Administration under Grant No. NNX08AR22G issued through the Planetary Science Division of the NASA Science Mission Directorate, the National Science Foundation under Grant No. AST-1238877, the University of Maryland, and Eotvos Lorand University (ELTE).

\textbf{IceCube:} The IceCube collaboration acknowledges the significant contributions to this manuscript from Anna Franckowiak, Claudio Kopper and Jakob van Santen. The IceCube collaboration gratefully acknowledge the support from the following agencies:
USA -- U.S. National Science Foundation-Office of Polar Programs,
U.S. National Science Foundation-Physics Division,
Wisconsin Alumni Research Foundation,
Center for High Throughput Computing (CHTC) at the University of Wisconsin-Madison,
Open Science Grid (OSG),
Extreme Science and Engineering Discovery Environment (XSEDE),
U.S. Department of Energy-National Energy Research Scientific Computing Center,
Particle astrophysics research computing center at the University of Maryland,
Institute for Cyber-Enabled Research at Michigan State University,
and Astroparticle physics computational facility at Marquette University;
Belgium -- Funds for Scientific Research (FRS-FNRS and FWO),
FWO Odysseus and Big Science programmes,
and Belgian Federal Science Policy Office (Belspo);
Germany -- Bundesministerium f\"ur Bildung und Forschung (BMBF),
Deutsche Forschungsgemeinschaft (DFG),
Helmholtz Alliance for Astroparticle Physics (HAP),
Initiative and Networking Fund of the Helmholtz Association,
Deutsches Elektronen Synchrotron (DESY),
and High Performance Computing cluster of the RWTH Aachen;
Sweden -- Swedish Research Council,
Swedish Polar Research Secretariat,
Swedish National Infrastructure for Computing (SNIC),
and Knut and Alice Wallenberg Foundation;
Australia -- Australian Research Council;
Canada -- Natural Sciences and Engineering Research Council of Canada,
Calcul Qu\'ebec, Compute Ontario, Canada Foundation for Innovation, WestGrid, and Compute Canada;
Denmark -- Villum Fonden, Danish National Research Foundation (DNRF), Carlsberg Foundation;
New Zealand -- Marsden Fund;
Japan -- Japan Society for Promotion of Science (JSPS)
and Institute for Global Prominent Research (IGPR) of Chiba University;
Korea -- National Research Foundation of Korea (NRF);
Switzerland -- Swiss National Science Foundation (SNSF).

\textbf{Others:} 

We thank the anonymous referee for useful comments.

EK, SJS and KS acknowledge support from the Science and Technology Facilities Council (STFC; ST/P000312/1, ST/N002520/1).
SJS acknowledges funding from the European Research Council under the European Union's Seventh Framework Programme (FP7/2007-2013)/ERC Grant agreement n$^{\rm o}$
[291222]. 

TWC acknowledgments the funding provided by the Alexander von Humboldt Foundation.

AF thanks Irene Tamborra for fruitful discussions on neutrino production in supernovae. AF is supported by the
Initiative and Networking Fund of the Helmholtz Association. 

This work is based (in part) on observations collected at the European Organisation for Astronomical Research in the Southern Hemisphere, Chile as part of PESSTO, (the Public ESO Spectroscopic Survey for Transient Objects Survey) ESO program 188.D-3003 and the 
DDT programme 297.D-5024.

Based on observations obtained at the Gemini Observatory, which is operated by the Association of Universities for Research in Astronomy, Inc., under a cooperative agreement with the NSF on behalf of the Gemini partnership: the National Science Foundation (United States), the National Research Council (Canada), CONICYT (Chile), Ministerio de Ciencia, Tecnolog\'{i}a e Innovaci\'{o}n Productiva (Argentina), and Minist\'{e}rio da Ci\^{e}ncia, Tecnologia e Inova\c{c}\~{a}o (Brazil). Observations were carried out under programme GN-2016A-DD-9.

Based on observations obtained with MegaPrime/MegaCam, a joint project of CFHT and CEA/IRFU, at the Canada-France-Hawaii Telescope (CFHT) which is operated by the National Research Council (NRC) of Canada, the Institut National des Science de l'Univers of the Centre National de la Recherche Scientifique (CNRS) of France, and the University of Hawaii. This work is based in part on data products produced at Terapix available at the Canadian Astronomy Data Centre as part of the Canada-France-Hawaii Telescope Legacy Survey, a collaborative project of NRC and CNRS. 

We have made use of the Weizmann interactive supernova data repository -- \url{http://wiserep.weizmann.ac.il}.

This research has made use of the NASA/IPAC Extragalactic Database (NED) which is operated by the Jet Propulsion Laboratory, California Institute of Technology, under contract with the National Aeronautics and Space Administration. 

Funding for SDSS-III has been provided by the Alfred P. Sloan Foundation, the Participating Institutions, the National Science Foundation, and the U.S. Department of Energy Office of Science. The SDSS-III web site is \url{http://www.sdss3.org/}.

SDSS-III is managed by the Astrophysical Research Consortium for the Participating Institutions of the SDSS-III Collaboration including the University of Arizona, the Brazilian Participation Group, Brookhaven National Laboratory, Carnegie Mellon University, University of Florida, the French Participation Group, the German Participation Group, Harvard University, the Instituto de Astrofisica de Canarias, the Michigan State/Notre Dame/JINA Participation Group, Johns Hopkins University, Lawrence Berkeley National Laboratory, Max Planck Institute for Astrophysics, Max Planck Institute for Extraterrestrial Physics, New Mexico State University, New York University, Ohio State University, Pennsylvania State University, University of Portsmouth, Princeton University, the Spanish Participation Group, University of Tokyo, University of Utah, Vanderbilt University, University of Virginia, University of Washington, and Yale University.

\end{acknowledgements}

\clearpage

\begin{appendix}
\section{Details of Pan-STARRS1 observations of other IceCube Events}
\label{sec:app}

In the tables that follow, we provide the information on the most probable type for each transient found: 
\begin{itemize}
    \item Stellar: the transient is spatially coincident with a faint source likely to be a star.
    \item QSO: the transient is spatially coincident with the core of a galaxy that has previously been catalogued as an AGN or QSO.
    \item SN: the transient is associated with a host galaxy at a plausible physical or angular separation.
    \item NT: nuclear transient. The transient is coincident with the core of a galaxy that has not been previously catalogued as an AGN or QSO. These could be AGN, or nuclear SNe or even tidal disruption events. 
    \item Orphan: there is no plausible host galaxy or stellar counterpart detected either in the Pan-STARRS reference images or any known catalogue.  
\end{itemize}

There were no transients that we found that had characteristics that made them peculiar from any others 
in the field and would lead to further investigation as
to their association with the neutrinos. 
Typically, the IceCube sky locations have a 90\% confidence region given by a likelihood scan with a radius of 0\fdg5$-$1\fdg2. However, these values depend on the sophistication of the IceCube track analysis. Therefore, for consistency and completeness, we report all transients in the tables below that were found within the Pan-STARRS1 camera footprint of around 1\fdg4 radius.

\subsection{IceCube-160427A}

The SN PS16cgx has been extensively discussed in the main manuscript. The rest of the transients discovered within the 90\% confidence region of IceCube-160427A are shortly summarised below.

PS16cfv is associated with SDSS J160133.99+094759.0. This is a stellar object in SDSS.

PS16cgw is offset from the host SDSS J160039.70+093919.6 by 1\farcs6\,N and 0\farcs2\,E.

PS16chy is associated with a stellar object SDSS J160044.22+095040.6.

PS16cfy is matched with SDSS J160205.21+094630.3, which is a spectroscopic QSO at $z = 2.344$.

PS16dow is offset from the host SDSS J160202.18+094907.6 core by 0\farcs1 which is Milliquas source at a photometric redshift of $z = 1.325$.

PS16dov is coincident with the spectroscopic QSO SDSS J160153.91+095244.3 at $z = 2.282$.

PS16cfu is offset by 0\farcs1\,N and 0\farcs2\,E from the host SDSS J160115.65+092504.8.

PS16cgb is coincident with a host SDSS J160219.12+093450.2 at a photometric redshift of $z = 0.755$. The transient light curve is flat; it could be an old SN or AGN variability.

PS16cfz is closely matched with the nucleus of SDSS J160211.95+095407.7 at a photometric redshift of $z = 0.169$. It is either a SN or AGN variability with a flat light curve.

PS16fns is coincident with SDSS J160124.77+100201.5; an AGN in Milliquas at a photometric redshift of $z = 1.955$.

PS16cfw is matched with the spectroscopic QSO SDSS J160139.09+092159.5 at $z = 0.862$.

PS16fnt is matched with the Milliquas AGN SDSS J160128.08+100250.9 at a photometric redshift $z = 0.975$.

PS16fnu is consistent with SDSS J155941.20+094619.4. This is a stellar object in SDSS.

PS16fnv is associated with SDSS J155941.54+094749.8. This is a source in the Milliquas catalogue at a photometric redshift of $z = 1.575$.

PS16fnw is coincident with SDSS J160153.37+100521.7, a spectroscopic QSO at $z = 2.285$.

PS16fnx is located 0\farcs3\,N and 0\farcs2\,W from SDSS J155928.45+095151.4. This is a faint galaxy in SDSS at a photometric redshift of $z = 0.579$. 

PS16fny host SDSS J160227.74+100309.0 is a source in Milliquas at a photometric redshift of $z = 0.375$.

PS16fnz is coincident with SDSS J160119.77+101218.7, which is a Milliquas source at a photometric redshift of $z = 0.225$.


\subsection{IceCube-160731A}

IceCube-160731A was discovered on 31 July 2016 01:55:04.00 UT (MJD=57600.07990). Pan-STARRS1 began observing the field 31 July 2016 07:41:39 UT (MJD=57600.32) and continued to observe every night until 7 August 2016, and then took another 6 epochs up until 26 August 2016. This gave an observing baseline of 26 days. Table \ref{tab:160731A} contains the candidate counterparts identified by the Pan-STARRS1 follow-up campaign, out 
to a 1\fdg45 radius. Three transient were spectroscopically classified. 

PS16dpm was discovered on the first epoch of observing and was observed to have a rising light curve. A spectrum was taken with the University of Hawaii SNIFS spectrometer on 3 August 2016, showing it to be a convincing Type Ia at $z = 0.12$ around 6 days before maximum light. A PESSTO spectrum confirmed this result \citep{smartt15,reynolds16}. The Pan-STARRS1 light curve and these two spectra imply a peak of MJD=57610 and an explosion epoch of MJD=57592$\pm$2, or 8 days before IceCube-160731A.

PS16dpl was also discovered on the first epoch of observing and the light curve stayed flat for 7 days in the
\ips\ band. The host galaxy SDSS J142303.54+000552.1 has a spectroscopic redshift of $z = 0.141$. It was observed by SNIFS on 7 August 2016 when it was still $m_{i_{\rm P1}}=20.8\pm0.1$ mag and shows features consistent with a Type Ia at this redshift.

PS16dpi was a known, old SN discovered by the SkyMapper transient survey as SMT16ojf \citep{scalzo17}, it was an old Type Ia at $z = 0.085115$ \citep{reynolds16} which exploded about 30 days before IceCube-160731A. 

The rest of the transients are summarised below and none had any characteristic to make it appear possibly related to IceCube-160731A. 

PS16cxe is not matched to any known or catalogued source.

PS16dpu has a SDSS host which is stellar.

6H3Ghhv is associated with SDSS J141721.29-003050.5. This is a faint and stellar object in SDSS.

6H3Ghhn is a marginal candidate, being faint and low signal-to-noise in \zps\ and absent in \ips. 

6H3Ghik is associated with SDSS J141718.86-003927.2 and is offset from the host core by 0\farcs2.

PS16dqg is not matched to any known or catalogued source.

6H3Ghhq is associated with SDSS J141843.68-005204.3. This is a faint and stellar object in SDSS.

PS16dpe is not matched to any known or catalogued source, but appears to be a SN at peak.

PS16dpc is not matched to any known or catalogued source, but appears to be a genuine orphaned SN.

PS16dpv is not matched to any known source and has a flat light curve.

PS16dpd is coincident with a stellar like source that was classified in SDSS DR12 as a star, but Milliquas catalogue \citep{flesch15} suggests it is a QSO. In any event, it is not a SN candidate.

PS16dpf is not matched to any known source and has a flat light curve. There appears to be a faint host in the Pan-STARRS1 reference image making this a potential SN candidate.

6H3Ghhx is associated with SDSS J141445.76-002739.7 but has a projected offset of only 0.3\,kpc, at the host redshift, implying it may be a nuclear transient.

PS16cyu is offset from SDSS J141953.14-010041.5 and shows a declining light curve. It was previously discovered by Pan-STARRS on 28 June 2016 and is therefore an old SN. 

6H3Ghhk is matched to SDSS J142101.10-004537.9. This is a faint and stellar object in SDSS.

PS16dpk, although not matched to a known source, showed a rising light curve after discovery marking it as a plausible SN candidate.

6H3Ghis is associated with SDSS J141506.06+001806.3, a stellar source at a photometric redshift of $z \sim 0.3$. 

PS16dpt has a SDSS host which is stellar.

6H3Ghhu is not matched to any known or catalogued source.

6H3Ghia is associated with SDSS J141417.75+000207.8 and is offset from the host by 0\farcs10\,N and 0\farcs20\,W

PS16dps is not matched to any known source, but does appear near an extended source in the Pan-STARRS1 footprint. This is a plausible SN candidate.

6H3Ghet is matched to SDSS J142152.68-004016.9 and is located 0\farcs03\,S and 0\farcs08\,W of the host core. At a redshift of 0.072, this gives a projected offset of approximately 10\,kpc.

6H3Ghjd is associated with SDSS J141600.56-011920.2 and is offset from the host by 0\farcs2, or a projected offset of 0.5\,kpc 
at a redshift of 0.171.

PS16dpj is not matched to any catalogued source and has a relatively flat light curve, making discerning the nature of this transient difficult.

6H3Ghib is associated with SDSS J141318.12-002527.8 and is offset from the host by 0\farcs16\,S and 0\farcs09\,W.

6H3Ghij is possibly associated with SDSS J141442.38+004325.7, a stellar source, or a QSO in Milliquas at a redshift of $z = 0.5$.

PS16dqf appears to be a nuclear transient.

PS16dpr is associated with a host catalogued in Milliquas, making it a likely QSO at a photometric redshift of $z = 2.043$.

\begin{table*}
\caption{31 transients discovered within the local vicinity of the IceCube-160731A neutrino detection (MJD 57600.07990). All transients reported appear within 1\fdg45 of the final best-fit neutrino localization.. All transient object magnitudes are reported in the \ips\ band unless otherwise noted. The separating line at 0\fdg75 indicates the transients within the Pan-STARRS1 footprint that are within the IceCube 90\% error location (see Table\,\ref{tab:summ}).
}
\begin{tabular}{lcrclcccc}
\hline
\hline
\multicolumn{1}{c}{Transient} & $\alpha_{\mathrm{J2000}}$ & \multicolumn{1}{c}{$\delta_{\mathrm{J2000}}$} & Disc. epoch & \multicolumn{1}{c}{$m_{\mathrm{disc},i_{\rm P1}}$} & Type & $z^{a}$ & $m_{\mathrm{host},i}^{b}$ &  $\Delta\Psi^{c}$\\
 & (hh:mm:ss.ss) & \multicolumn{1}{c}{(dd:mm:ss.s)} & (MJD) & \multicolumn{1}{c}{(mag)} & & & (mag) & ($^{\circ}$) \\
\hline
PS16cxe & 14:18:00.02 & -00:14:35.3 & 57600.56667 & 21.11 & Orphan & $-$ & $-$ & 0.06 \\
PS16dpu & 14:17:43.93 & -00:33:28.5 & 57602.57778 & 21.62$^{d}$ & Stellar &	$-$	& 21.49 & 0.27 \\
6H3Ghhv & 14:17:21.30 & -00:30:50.4 & 57600.56667 & 22.26 & Old SN & $-$ & 21.48 & 0.27 \\
6H3Ghhn & 14:17:40.06 & -00:37:59.9 & 57601.57500 & 20.84$^{d}$ & NT & 0.19801$\pm$0.00004$^{e}$ & 17.31 & 0.34 \\
6H3Ghik & 14:17:18.89 & -00:39:27.2 & 57601.57500 & 21.37$^{d}$ & NT & 0.143$\pm$0.026 & 18.00 & 0.40 \\
PS16dqg & 14:16:02.51 & -00:02:55.1 & 57603.59792 & 22.15$^{d}$ & Old SN & 0.60742$\pm$0.00016$^{e}$ & 19.71 & 0.55 \\
6H3Ghhq & 14:18:31.87 & -00:50:39.0 & 57601.57500 & 22.09$^{d}$ & Old SN & $-$ & 22.64 & 0.56 \\
PS16dpe & 14:17:12.43 & +00:17:23.0 & 57600.56667 & 22.04 & Old SN & 0.293$\pm$0.103 & 19.89 & 0.62 \\
PS16dpc & 14:15:14.83 & -00:21:15.7 & 57600.56667 & 21.58 & Orphan & $-$ & $-$ & 0.69 \\
PS16dpv & 14:20:13.30 & -00:45:52.3 & 57601.57500 & 22.51 & Old SN & $-$ & 23.58 & 0.72 \\
PS16dpd & 14:16:31.15 & -00:56:56.2 & 57600.56667 & 20.67 & Likely QSO & 1.300$^{f}$ & 20.54 & 0.75 \\
\hline
PS16dpf & 14:18:32.97 & +00:30:40.7 & 57600.56667 & 21.86 & Old SN & 0.679$\pm$0.140 & 21.00 & 0.82 \\
6H3Ghhx & 14:14:45.75 & -00:27:39.9 & 57601.57500 & 21.24$^{d}$ & NT & 0.159$\pm$0.023 & 17.45 & 0.83 \\
PS16cyu & 14:19:53.22 & -01:00:39.6 & 57600.56667 & 20.97 & Old SN & 0.135$\pm$0.034 & 18.53 & 0.85 \\
6H3Ghhk & 14:21:01.11 & -00:45:38.0 & 57601.57500 & 22.00$^{d}$ & Old SN & $-$ & 21.92 & 0.88 \\
PS16dpk & 14:21:22.51 & -00:47:02.9 & 57600.56667 & 22.00 & Stellar & $-$ & 22.51 & 0.97 \\
6H3Ghis & 14:15:06.08 & +00:18:05.9 & 57602.57708 & 21.26$^{d}$ & NT & 0.319$\pm$0.053 & 19.42 & 0.94 \\
PS16dpt & 14:17:19.09 & +00:37:43.2 & 57602.57778 & 22.07 & Stellar & $-$ & 21.90 & 0.94 \\
6H3Ghhu & 14:16:31.86 & +00:34:52.8 & 57602.57778 & 22.28$^{d}$ & Orphan & $-$ & $-$ & 0.95 \\
6H3Ghia & 14:14:17.77 & +00:02:07.8 & 57601.57500 & 21.51$^{d}$ & Old SN & 0.424$\pm$0.055 & 20.49 & 0.98 \\
PS16dps & 14:13:57.45 & -00:10:43.4 & 57600.56667 & 21.96 & Old SN & 0.152$\pm$0.074 & 19.92 & 1.02 \\
6H3Ghet & 14:21:52.69 & -00:40:16.9 & 57601.57500 & 22.04$^{d}$ & Old SN & 0.07247$\pm$0.00003$^{e}$ & 17.33 & 1.04 \\
PS16dpi & 14:21:13.16 & +00:25:41.6 & 57600.56667 & 19.26 & SN Ia &	0.085115$\pm$0.000297$^{g}$ & 16.52 & 1.09 \\
6H3Ghjd & 14:16:00.58 & -01:19:20.4 & 57602.57708 & 21.30$^{d}$ & NT & 0.17135$\pm$0.00003$^{e}$ & 17.10 & 1.14 \\
PS16dpj & 14:21:16.17 & -01:05:54.4 & 57600.56667 & 21.39 & Old SN & 0.141$\pm$0.116 & 21.49 & 1.14 \\
6H3Ghib & 14:13:18.14 & -00:25:27.6 & 57600.56667 & 21.26 & Old SN & 0.542$\pm$0.154 & 19.95 & 1.18 \\
6H3Ghij & 14:14:42.39 & +00:43:25.7 & 57600.56667 & 21.65 & Likely QSO & 0.5$^{f}$ & 20.07 & 1.31 \\
PS16dpl & 14:23:03.19 & +00:05:54.5 & 57600.56667 & 20.83 & SN Ia &	0.14129$\pm$0.00002$^{e}$ & 18.21 & 1.32 \\
PS16dpm & 14:23:20.15 & -00:32:16.5 & 57600.56667 & 20.35 & SN Ia &	0.12$^{h}$ & 21.01 & 1.35 \\
PS16dqf & 14:14:37.64 & +00:46:05.8 & 57603.59792 & 21.68$^{d}$ & Old SN & 0.464$\pm$0.078 & 20.35 & 1.36 \\
PS16dpr & 14:12:29.06 & -00:36:12.7 & 57600.56667 & 20.41 & Likely QSO & 2.043$^{f}$ & 19.66 & 1.41 \\
\hline
\label{tab:160731A}
\end{tabular}
\begin{flushleft}
$^{a}$ Redshifts presented are photometric redshifts to host galaxies taken from SDSS DR12 unless otherwise noted.\\
$^{b}$ Host magnitudes presented are taken from SDSS DR12.\\
$^{c}$ Angular distance to the final revised coordinates of IceCube-160731A.\\  
$^{d}$ Magnitudes are from discovery in the \zps\ band.\\
$^{e}$ Spectroscopic redshift to host galaxy taken from SDSS DR12.\\
$^{f}$ Photometric redshift from Milliquas.\\
$^{g}$ Spectroscopic redshift to host galaxy taken from NED.\\
$^{h}$ Spectroscopic redshift of the SN.
\end{flushleft}
\end{table*}

\subsection{IceCube-161103A}

IceCube-161103A was discovered on 3 November 2016 09:07:31.12 (MJD 57695.38022). Pan-STARRS1 follow-up observations began the same night on 3 November 2016 11:09:55 UT (MJD 57695.47) and proceeded for a subsequent 9 nights. Of these, 2 nights were lost due to weather. A total of 7 transients were found while searching the confidence region in this time, detailed in Table \ref{tab:161103A}. None of these events were spectroscopically classified.

PS16exc is not associated with any known or catalogued source and is faint, at $\sim$22 mag in \ips. The light curve remains relatively flat for almost 25 days.

PS16exd is not associated with any known or catalogued sources. The transient was observed for several days in the \ips\ band before fading and it is a potential SN candidate at maximum. The host is a faint object, either a compact galaxy or star. 

PS16exh has no associated catalogued host but a faint point source is visible in the the Pan-STARRS1 reference stacks. This is a potential stellar source but appears more as a variable QSO. 

PS16exi has no associated catalogued host but a faint point source is visible in the the Pan-STARRS1 reference stacks. This is a potential stellar source but appears more as a variable QSO. 

PS16exf  has no associated catalogued host but a faint point source is visible in the the Pan-STARRS1 reference stacks. This is likely a stellar source or variable QSO. 
 
PS16exg is not associated with any known or catalogued sources. The Pan-STARRS1 exposures show a point source, coupled with a relatively flat light curve (declining $<$0.5 mag in 44 days) suggests this may be an AGN/QSO or a stellar source. 

6K3Ghjo is not associated with any known or catalogued source. It was observed initially in \ips\ but also in \zps\ from +40 days from discovery. Typically, these \zps\ band observations were one magnitude brighter than the \ips\ band data, but both sets were relatively flat, implying a stellar source. There is a clear point source at the transient position. 

PS16fhi was found outside the 90\% radius (at $\alpha_{\mathrm{J2000}}$ = 02$^{\mathrm{h}}$38$^{\mathrm{m}}$35$\fs$74, $\delta_{\mathrm{J2000}}$ = +13$^{\circ}$05\arcmin06\farcs2) and is not listed in in Table\,\ref{tab:161103A}. However it initially caught our attention as it began quickly rising. A Gemini GMOS spectrum and its rising lightcuve showed it to be a bright Type II at $z = 0.085$, resembling SN~2009kf \citep{botticella10}. We rule it out as being related as a combination of these data implied it exploded significantly after (up to 10$-$12 days) IceCube-161103A.

\begin{table*}
\caption{7 transients discovered within the local vicinity of the IceCube-161103A neutrino detection (MJD 57695.38022). All transients reported appear within 1\fdg1 of the final best-fit neutrino localization.}
\begin{tabular}{lcccccccc}
\hline
\hline
\multicolumn{1}{c}{Transient} & $\alpha_{\mathrm{J2000}}$ & $\delta_{\mathrm{J2000}}$ & Disc. epoch & $m_{\mathrm{disc},i_{\rm P1}}$ & Type & $z$ & $m_{\mathrm{host},i_{\rm P1}}$ &  $\Delta\Psi^{a}$\\
 & (hh:mm:ss.ss) & (dd:mm:ss.s) & (MJD) & (mag) & & & (mag) & ($^{\circ}$) \\
\hline
PS16exc & 02:41:05.69 & +12:55:59.3 & 57695.46875 & 21.52 & Orphan & $-$ & $-$ & 0.61 \\
PS16exd & 02:41:08.05 & +13:03:59.2 & 57695.48961 & 21.81 & Old SN & $-$ & 21.21 & 0.69 \\
PS16exh & 02:44:56.02 & +13:15:50.7 & 57695.46875 & 21.25 & Likely QSO & $-$ & 21.57 & 0.79 \\
PS16exi & 02:45:02.51 & +11:52:00.5 & 57695.48961 & 22.30 & Likely QSO & $-$ & 21.55 & 0.86 \\
PS16exf & 02:44:02.89 & +13:30:45.7 & 57695.47489 & 21.61 & Likely QSO & $-$ & 21.37 & 0.94 \\
PS16exg & 02:44:14.42 & +11:37:18.7 & 57695.47489 & 21.70 & Likely QSO & $-$ & 21.31 & 1.01 \\
6K3Ghjo & 02:47:45.09 & +12:46:20.8 & 57695.48961 & 22.66 & Stellar & $-$ & 21.10 & 1.12 \\
\hline
\label{tab:161103A}
\end{tabular}
\begin{flushleft}
$^{a}$ Angular distance to the final revised coordinates of IceCube-161103A.\\
\end{flushleft}
\end{table*}

\subsection{IceCube-170321A}

IceCube-170321A was discovered on 21 March 2017 07:32:20.69 UT (MJD 57833.31413). Pan-STARRS1 began following this event the next night on 22 March 2017 05:44:56 UT (MJD 57834.24) and proceeded for a subsequent 3 nights. From the night of 27 March 2017 to 31 March 2017, observations were performed on alternate nights. Following this, a final night was dedicated to observing on 6 April 2017. Table \ref{tab:170321A} contains the candidate counterparts identified by the Pan-STARRS1 follow-up campaign while searching the confidence region. A total of 7 transients were identified, though none were observed spectroscopically.

PS17ceb is not matched to any known or catalogued source, but does appear to be within a faint, extended source. This is most likely a galaxy, making this transient a SN candidate.

PS17cdy appears as a point source but is not matched to any known source. It is likely an uncatalogued stellar source.

PS17cdz is not matched to any known or catalogued source.

7C3Gjfw is not matched to any known or catalogued sources.

PS17cea is not matched to any known or catalogued source.

PS17byo is not matched to any known sources, but does show a faint potential host in the reference images. As such, this is likely a SN candidate.

PS17byn is not matched to any known or catalogued source, but its light curve shows evidence of some decline. It is possible that this is a SN candidate.

\begin{table*}
\caption{7 transients discovered within the local vicinity of the IceCube-170321A neutrino detection (MJD 57805.31413). The separating line at 1\fdg2 indicates the transients within the Pan-STARRS1 footprint that are within the IceCube 90\% error location (see Table\,\ref{tab:summ}).
}
\begin{tabular}{lcccccccc}
\hline
\hline
\multicolumn{1}{c}{Transient} & $\alpha_{\mathrm{J2000}}$ & $\delta_{\mathrm{J2000}}$ & Disc. epoch & $m_{\mathrm{disc},i_{\rm P1}}$ & Type & $z$ & $m_{\mathrm{host},i_{\rm P1}}$ &  $\Delta\Psi^{a}$\\
 & (hh:mm:ss.ss) & (dd:mm:ss.s) & (MJD) & (mag) & & & (mag) & ($^{\circ}$) \\
\hline
PS17ceb	& 06:36:32.74 & -15:15:51.9 & 57834.27474 & 20.90 & Old SN & $-$ & 19.18 & 0.85 \\
PS17cdy	& 06:29:00.76 & -15:27:28.7 & 57834.24305 & 20.55 & Stellar & $-$ &	20.13 & 1.11 \\
\hline
PS17cdz	& 06:27:59.29 & -15:09:06.1 & 57843.26768 & 20.58 & Orphan & $-$ & $-$ & 1.27 \\
7C3Gjfw	& 06:37:41.71 & -14:17:43.7 & 57834.26376 & 21.99 & Orphan & $-$ & $-$	& 1.29 \\
PS17cea	& 06:34:43.80 & -13:41:48.8 & 57834.26376 & 22.29 & Orphan & $-$ & $-$	& 1.35 \\
PS17byo	& 06:27:54.74 & -14:26:07.5 & 57834.24922 & 21.61 & Old SN & $-$ & 21.86 & 1.40 \\
PS17byn	& 06:27:58.43 & -15:59:18.9 & 57834.25582 & 21.42 & Orphan & $-$ & $-$ & 1.60 \\
\hline
\label{tab:170321A}
\end{tabular}
\begin{flushleft}
$^{a}$ Angular distance to the final revised coordinates of IceCube-170321A.\\
\end{flushleft}
\end{table*}

\subsection{IceCube-171106A}

The field of IceCube-171106A was observed starting from 8 November 2017 08:09:36.0 UT (MJD 58065.34), some 37.44 hours following the detection and proceeded for the subsequent 10 nights. Observations were made using the \ips\ and \zps\ bands, with difference imaging being performed with respect to the Pan-STARRS1 3$\pi$ stacked reference sky. 5$\sigma$ limiting magnitudes of $m_{i_{\rm{P1}}} \approx 22.5$ mag were achieved. Table \ref{tab:171106A} contains the candidate counterparts identified by the Pan-STARRS1 follow-up campaign. The initial results of the IceCube-160427A Pan-STARRS1 follow-up were reported by \citet{mcbrien17}.

PS17eym is a SN candidate associated with SDSS J223922.20+071638.0. It is located 1\farcs80\,S and 0\farcs97\,W from the galaxy core, or a projected offset of 10.7\,kpc.

PS17fcc is a SN candidate that was found associated with the galaxy SDSS J224158.66+074653.2, being located 0\farcs59\,N and 0\farcs28\,W from the galaxy core, or a projected offset of 1.9\,kpc.

PS17eyn is a SN candidate associated with the galaxy SDSS J224224.58+065502.1. It is located 0\farcs58\,N and 0\farcs84\,E from the galaxy core, or a projected offset of 1.8\,kpc.

PS17fem is a SN candidate associated with the galaxy SDSS J224219.29+064237.5. It is offset from the galaxy core by 0\farcs57\,N and 0\farcs41\,E or a projected offset of 2.1\,kpc.

PS17eyo is a SN candidate that is not associated with any known or catalogued host.

PS17eyl is a nuclear transient, being associated with the galaxy SDSS J223902.90+062526.6. It is offset from the galaxy core by 0\farcs28\,S and 0\farcs04\,E, a projected offset of 1.1\,kpc.

PS17fcd is a nuclear transient located in the galaxy SDSS J224401.85+081327.5, 0\farcs37\,N and 0\farcs13\,W from the core. This is a projected offset of 2.1\,kpc. This transient was first discovered in the \zps\ band on MJD 58066.20450 at magnitude 20.53.

\begin{table*}
\caption{7 transients discovered within the local vicinity of the IceCube-171106A neutrino detection (MJD 58063.77754). All transient object magnitudes are reported in the \ips\ band unless otherwise noted.
The separating line at 0\fdg7 indicates the transients within the Pan-STARRS1 footprint that are within the IceCube 90\% error location (see Table\,\ref{tab:summ}).
}
\begin{tabular}{lccclcccc}
\hline
\hline
\multicolumn{1}{c}{Transient} & $\alpha_{\mathrm{J2000}}$ & $\delta_{\mathrm{J2000}}$ & Disc. epoch & \multicolumn{1}{c}{$m_{\mathrm{disc},i_{\rm P1}}$} & Type & $z^{a}$ & $m_{\mathrm{host},i}^{b}$ &  $\Delta\Psi^{c}$\\
 & (hh:mm:ss.ss) & (dd:mm:ss.s) & (MJD) & \multicolumn{1}{c}{(mag)} & & & (mag) & ($^{\circ}$) \\
\hline
PS17eym & 22:39:22.13 & +07:16:36.4 & 58065.37269 & 21.48 & Old SN & 0.380$\pm$0.106 & 20.14 & 0.20 \\
PS17fcc & 22:41:58.65 & +07:46:53.6 & 58065.37269 & 21.54 & Old SN & 0.172$\pm$0.032 & 19.04 & 0.62 \\
\hline
PS17eyn & 22:42:24.63 & +06:55:02.7 & 58065.34415 & 20.62 & Old SN & 0.095$\pm$0.022 & 17.23 & 0.77 \\
PS17fem & 22:42:19.29 & +06:42:38.0 & 58066.25399 & 21.95 & Old SN & 0.178$\pm$0.016 & 17.79 & 0.90 \\
PS17eyo & 22:40:40.92 & +06:26:29.6 & 58065.37269 & 22.16 & Orphan & $-$ & $-$ & 0.97 \\
PS17eyl & 22:39:02.90 & +06:25:26.3 & 58065.37269 & 21.23 & Old SN & 0.237$\pm$0.095 & 22.14 & 1.00 \\
PS17fcd & 22:44:01.84 & +08:13:27.6 & 58066.20450 & 20.53$^{d}$ & Old SN & 0.403$\pm$0.092 & 19.62 & 1.30 \\
\hline
\label{tab:171106A}
\end{tabular}
\begin{flushleft}
$^{a}$ Redshifts presented are photometric redshifts to host galaxies taken from SDSS DR12.\\
$^{b}$ Host magnitudes presented are taken from SDSS DR12.\\
$^{c}$ Angular distance to the final revised coordinates of IceCube-171106A.\\
$^{d}$ Discovered in the \zps\ band. The earliest \ips\ band observation was on MJD 58067.22367 at magnitude 21.15.
\end{flushleft}
\end{table*}

\end{appendix}

\end{document}